\documentclass[a4paper,fleqn,usenatbib]{mnras}

\usepackage{newtxtext,newtxmath}

\usepackage[T1]{fontenc}
\usepackage{ae,aecompl}

\newcommand\msun{\, \rm M_\odot}
\newcommand\pc{{\, \rm pc}}
\newcommand\mbin{{M_{\rm b}}}
\newcommand\mgal{{M_{\rm g}}}

\usepackage{graphicx}	
\usepackage{amsmath}	
\usepackage{amssymb}	

\usepackage{BibDef}
\usepackage{geometry}
\usepackage{lipsum}

\usepackage{color}

\title[Influence of MBH Binaries on Merger Remnants]{The influence of Massive Black Hole Binaries on the Morphology of Merger Remnants}

\author[E. Bortolas et al.]{
E. Bortolas$^{1,2}$\thanks{E-mail: elisa.bortolas@oapd.inaf.it},
A. Gualandris$^{3}$,
M. Dotti$^{4,5}$ and J. I. Read$^3$
\\
$^{1}$INAF, Osservatorio Astronomico di Padova, Vicolo dell'Osservatorio 5, 35122, Padova, Italy\\
$^{2}$Dipartimento di Fisica e Astronomia `Galileo Galilei', Universit\`a di Padova, Vicolo dell'Osservatorio 3, 35122 Padova, Italy\\
$^{3}$Department of Physics, University of Surrey, Guildford GU2 7XH, United Kingdom\\
$^{4}$Dipartimento di Fisica G. Occhialini, Universit\`a degli Studi di Milano, Bicocca, Piazza della Scienza 3, 20126 Milano, Italy\\
$^{5}$INFN, Sezione di Milano-Bicocca, Piazza della Scienza 3, 20126 Milano, Italy\\
}

\date{Accepted XXX. Received YYY; in original from ZZZ}

\pubyear{2017}

\begin{document}
\label{firstpage}
\pagerange{\pageref{firstpage}--\pageref{lastpage}}
\maketitle

\begin{abstract}
Massive black hole (MBH) binaries, formed as a result of galaxy mergers, are expected to harden by dynamical friction and  three-body stellar scatterings, until emission of gravitational waves (GWs) leads to their final coalescence. According to recent simulations, MBH binaries can efficiently harden via stellar encounters only when the host geometry is triaxial, even if only modestly, as angular momentum diffusion allows an efficient repopulation of the binary loss cone. In this paper, we carry out a suite of $N$-body simulations of equal-mass galaxy collisions, varying the initial orbits and density profiles for the merging galaxies and running simulations both with and without central MBHs. We find that the presence of an MBH binary in the remnant makes the system 
 nearly oblate, {  aligned with the galaxy merger plane}, within a radius enclosing 100 MBH masses. We never find binary hosts to be  prolate on any scale. The decaying MBHs slightly enhance the tangential anisotropy in the centre of the remnant due to angular momentum injection and the slingshot ejection of stars on nearly radial orbits. This latter effect results in about 1\% of the remnant stars being expelled from the galactic nucleus. Finally, we do not find any strong connection between the remnant morphology and the binary hardening rate, which depends only on the inner density slope of the remnant galaxy. Our results suggest that MBH binaries are able to coalesce within a few Gyr, even if the binary is found to partially erase the merger-induced triaxiality from the remnant.
\end{abstract}




\section{Introduction}\label{sec:intro}
In the standard cosmological model, galaxies grow through the successive mergers of smaller galaxies \citep[e.g.][]{WhiteRees1978}. Combined with the observational evidence that massive black holes (MBHs) dwell in galaxy centres from early times, this suggests that a large number of massive black hole binaries (BHBs) must have formed over cosmic time  \citep{Haehnelt1993,Wu2015}. After the galactic collision, BHBs  reduce their separation via 
dynamical friction and slingshot ejections of stars on 
intersecting orbits \citep{Saslaw1974,Begelman1980}. If the hardening continues down to 
separations of a few milliparsecs, BHBs are expected to reach 
coalescence in a burst of gravitational waves \citep[GWs, ][]{Thorne1976}; as such, they represent one of the most powerful sources of
GWs in the low frequency range accessible to the Pulsar Timing Array
 and future space-based observatories like LISA \citep[e.g.][]{Hobbs2010,Babak2016,Amaro-Seoane2017}. 
The detection of such signals would give unprecedented 
information on MBH properties and allow to test
the current cosmological paradigm \citep{Hogan2009}. 

The late BHB evolution in gas poor environments 
 has been put under scrutiny
over the last decades: in the beginning of the slingshot phase, 
the BHB promptly expels most of the stars that are initially on  loss cone orbits due to a three-body encounter with the MBHs, thus  BHB 
hardening can persist only if the reservoir of stars on
low angular momentum orbits is readily 
replenished. In fact,  several  studies 
of BHBs hardening in spherical stellar environments have 
shown that the binary cannot shrink below $\sim$1 pc scale, 
as no efficient mechanism can guarantee a steady loss cone 
repopulation in spherical nuclei   \citep{Begelman1980,Milosavljevic2001,Yu2002,Makino2004}. The so-called \emph{final parsec problem} 
seems to prevent BHBs from merging within a Hubble time \citep{Milosavljevic2003b}.

Several mechanisms have been proposed as possible solutions, among them the influence of gas drag on the BHB orbital evolution \citep[e.g.][]{Escala2004, Dotti2007, Tang2017}  and the loss cone repopulation produced by the presence of a  massive perturber such  as a molecular cloud \citep[][Bortolas et al., in prep.]{Perets2008, Goicovic2017} or a  stellar cluster \citep{Bortolas2018,ArcaSedda2017}. The BHB random walk has also been suggested as a possible  booster for the loss cone refilling \citep{Milosavljevic2001, Chatterjee2003, Milosavljevic2003b}, but recent studies suggest that this effect is not relevant if the host system harbours more than $\sim 10^6$ stars \citep{Bortolas2016}.


Recently, a series of theoretical and numerical studies of BHBs hardening in different stellar environments  pinpointed a more general solution for the final parsec problem: in fact, the slingshot driven BHB hardening has been found to crucially depend on the shape of the merger remnant   \citep{Yu2002, MerrittPoon2004b, Berczik2006, Khan2013,Vasiliev2014}.  Such studies show that stars  are 
continually supplied to the BHB when the host geometry 
is triaxial, even if only modestly, as diffusion in angular momentum 
allows for efficient loss cone refilling even when two-body relaxation is negligible 
 \citep{Yu2002,Vasiliev2015,Gualandris2017}.
Departures from spherical symmetry are expected and are in fact observed in all merger remnants, implying that BHBs are able to reach final coalescence within a Hubble time in most galaxies \citep[e.g.][]{Khan2011, Preto2011,Gualandris2012,Khan2016}. Purely axisymmetric remnants, however, seem unable to drive BHBs to coalescence \citep{Vasiliev2015,Gualandris2017}.

Even if all merger remnants show some degree of asphericity \citep[e.g.][]{deZeeuw1991}, the actual shape of the relic is known to depend on several factors, primarily the initial orbit and the properties of the progenitors. Linking the morphology and kinematics of present-day galaxies to their formation and merger histories is a long standing challenge, dating back to the very first astrophysical simulations \citep[e.g. ][]{Holmberg1941,Toomre1972,White1978, White1979}. Galaxy mergers seem to play a major role in shaping present day ellipticals and in determining their size evolution \citep[e.g.][]{Cox2006,Naab2009,Oser2012,Hilz2012,Frigo2016}; a plethora of studies focus on the formation of ellipticals via mergers of spiral galaxies \citep[][and references therein]{Naab2013,Naab2016}, but the possibility of producing ellipticals via collisions of pressure-supported systems has also been explored\footnote{ 
It has long been known that present-day ellipticals cannot be formed from the mergers of present-day spirals \citep[e.g.][]{Ostriker1980,Cox2006}. 
Indeed, many ellipticals require dissipational mergers (i.e. mergers that bring in fresh gas and promote star formation) to produce their observed kinematics \citep[e.g][]{Dubinski1998, Hilz2013}. The most massive ellipticals, however, appear to only grow through dissipationless mergers \citep[e.g][]{Dubinski1998, Hilz2013}  and so our focus in this paper is on gas-free `dry' mergers.
}  \citep{White1978, White1979, Gonzalez2005, Gonzalez2005b, DiMatteo2009, Hilz2012}. An interesting result in such scenario is that 
the merging nuclei often experience a nearly head-on collision producing maximally triaxial or nearly prolate (bullet-like) remnants \citep{Gonzalez2005b,DiMatteo2009}. 

However, these findings may not apply if the colliding systems host an MBH. Early studies on the stability of triaxial systems including a single central MBH have shown that the massive body acts as a scattering centre, driving the system toward an oblate (disc-like) configuration \citep{Gerhard1985,MerrittQuinlan1998}. More recently,  \citet{MerrittPoon2001,MerrittPoon2002,MerrittPoon2004
} demonstrated the existence of equilibrium configurations  for maximally triaxial and nearly oblate systems hosting an MBH, even when a large fraction of chaotic orbits are included; however nearly prolate shapes seem not to be sustainable in the presence of a central MBH \citep{MerrittPoon2004}.

The above work suggests that the morphology and kinematics of dry-merger remnants will be altered if at least one MBH takes part in the galactic collision. This is important because, as explained above, the shape of the remnant has a strong influence on the hardening efficiency of any post-merger BHB \citep[e.g.][]{Gualandris2017}. 

In this paper, we explore the consequences of the presence of  BHBs on the geometry of their host galaxies for the first time. We start our simulations from the merger of two spherical stellar systems, and we study the evolution of the  remnant geometry when the BHB is not present and when it is included. We find that the central massive bodies strongly change the morphology of their hosts well beyond their  sphere of influence, leading the system towards oblate (disky) shapes. The study of the effect of the environment on the BHB hardening is of utmost importance, since it has  implications for the BHB coalescence rates expected for forthcoming GW observatories.

The paper is organized as follows: Section~\ref{sec:methods} presents the numerical methods and initial conditions of the simulations, while Section~\ref{sec:general}  lays out some useful theoretical concepts; Section~\ref{sec:res} presents the results of our simulations; finally, in Section~\ref{sec:disc} we present a summary and discussion.

\section{Methods}\label{sec:methods}

\subsection{Initial Conditions}
We consider the merger of two equal-mass  galaxies, both set up with an isotropic, spherically symmetric, Dehnen density profile \citep{Dehnen1993}:
\begin{equation}
\rho(r)=\frac{(3-\gamma) \,\mgal}{4\pi}\frac{r_0}{r^\gamma(r+r_0)^{4-\gamma}},
\end{equation}
where $\mgal$ is the total mass of the galaxy, $r_0$ is the scale radius of the model and $\gamma$ is its inner density slope. Each galaxy was sampled with $N=512k$ equal mass particles (we discuss our choice of force softening in section \ref{sec:simuls}). The merging galaxies were initially on a bound Keplerian orbit with  semimajor axis $a_i = 15r_0$ and separated by $\Delta r = 20r_0$. 

We ran different simulations changing the density slope of the merging galaxies and their initial orbital eccentricity. Specifically, we varied the density profile by setting $\gamma=0.5$ (low concentration systems, LC), $\gamma=1$ (medium concentration systems, MC) and $\gamma=1.5$ (high concentration systems, HC); we set the orbital eccentricity as  $e=0.5$ (runs labelled wit `5'), $e=0.7$ (runs labelled wit `7') and $e=0.9$ (runs labelled with `9') for a total of nine different configurations. We ran all the simulations both omitting and including  (runs labelled with `b') a MBH in the centre of each colliding system; this last case leads to the formation of a BHB in the centre of the remnant. When present, the mass of the MBH is $M_{\bullet} = 0.005 \mgal$. Table~\ref{tab:runs} lists the identifiers of each run in the suite of simulations.

{
We add a further run (MC7o) including a MBH in only one of the two colliding systems; the properties of the merging galaxies in this last simulation are the same as in MC7, MC7b, while the MBH mass is $M_{\bullet} = 0.01 \mgal$. }

\begin{table}
  \centering
  \caption{Identifiers of the runs. Columns refer to the  initial concentration of the merging galaxies ($\gamma=0.5, 1, 1.5$, respectively Low, Medium and High Concentration); rows refer to the initial orbital eccentricity ($e=0.5, 0.7, 0.9$; labels 5, 7, 9 respectively) and to whether { the merger remnant hosts no MBHs (no additional label), only one MBH (run labelled with `o') or a BHB (runs labelled with `b'). }}
  \label{tab:runs}
  \begin{tabular}{lccc} 
    \hline
                 & $\gamma=0.5$ (LC) & $\gamma=1$ (MC) & $\gamma=1.5$ (HC) \\
    \hline
     $e=0.5$, no MBHs    &  LC5              &  MC5             &  HC5 \\
     $e=0.7$, no MBHs    &  LC7              &  MC7             &  HC7 \\
     $e=0.9$, no MBHs    &  LC9              &  MC9             &  HC9  \\
     $e=0.5$, {BHB}      &  LC5b             &  MC5b            &  HC5b \\
     $e=0.7$, {BHB}      &  LC7b             &  MC7b            &  HC7b \\
     $e=0.9$, {BHB}      &  LC9b             &  MC9b            &  HC9b \\
     $e=0.7$, { one MBH}  &  -                &  MC7o            &  -    \\
    \hline
  \end{tabular}
\end{table}

\subsection{N-body Units}

\begin{table}
  \centering
  \caption{Scaling of the models depending on the inner density slope of the primordial galaxies ($\gamma$). $M_\bullet$ is the MBH mass at the centre of the Dehnen model, [M] is the unit mass, [L] is the length unit, [T] is the time unit and [V] is the velocity unit. }
  \label{tab:nbody}
  \begin{tabular}{lcccc} 
    \hline
            &   $M_\bullet$, [M] ($\msun$) & [L] (pc) & [T] (Myr) & [V] (km/s) \\
    \hline
$\gamma=0.5$&$4\times10^6$, $8\times10^8$& 30& $8.67\times10^{-2}$& 339 \\
$\gamma=1$  &$4\times10^6$, $8\times10^8$& 50& $1.86\times10^{-1}$& 262 \\
$\gamma=1.5$&$4\times10^6$, $8\times10^8$&120& $6.93\times10^{-1}$& 169 \\
\hline
$\gamma=0.5$&$10^8$, $2\times 10^{10}$   &190& $2.76\times10^{-1}$& 673 \\
$\gamma=1$  &$10^8$, $2\times 10^{10}$   &320& $6.04\times10^{-1}$& 518 \\
$\gamma=1.5$&$10^8$, $2\times 10^{10}$   &720& $2.03             $& 346 \\
    \hline
  \end{tabular}
\end{table}

Non-dimensional units are used throughout the paper: the Newtonian gravitational constant $G$ is set equal to 1, and we further set $r_0=M_{\rm tot}=1$, where $M_{\rm tot}$ is the total stellar mass  in each simulation (i.e. since we have equal mass mergers, $M_{\rm tot}=2 \mgal$). It is possible to rescale the systems to real galaxies when the MBHs are present by using the relation between the MBH influence radius ($r_{\rm infl}$, i.e. the radius including $2M_\bullet$ in stars) and the MBH mass ($M_\bullet$) using the relation presented in \citet{Merritt2009}: $r_{\rm infl} = 30 \pc \times (M_\bullet /10^8 \msun)^{0.56}$. Table~\ref{tab:nbody} lists the scaling units; $r_{\rm infl}$ is computed analytically as the radius enclosing a stellar mass equal to $2M_\bullet$ in the Dehnen profile considered. The same scaling is assumed for equivalent runs without MBHs.

\subsection{Simulations}\label{sec:simuls}

\begin{figure}
\center
\includegraphics[trim={1cm 0.35cm 7cm 20cm},width=0.95\columnwidth]{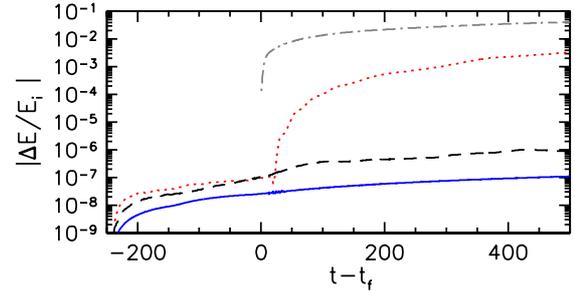}
 \caption{ The plot shows the time evolution of the relative energy error ${|\Delta E/E|}$ as defined in equation (\ref{eq:energy}) for the runs with $\gamma=1$ and $e=0.7$. In particular, we show the energy error for runs with no MBHs (solid blue line, run MC7), with only one MBH (dashed black line, run MC7o), and with a BHB (dotted red line, run MC7b). For comparison, we also show the BHB binding energy in run MC7b with a grey dash-dotted line; such quantity is always at least one order of magnitude larger than the relative global energy error in run MC7b.  }  \label{fig:en}
\end{figure}

The simulations are performed adopting the direct summation $N$-body code \texttt{HiGPUs} \citep{Capuzzo-Dolcetta2013}, designed to run on GPU accelerators. \texttt{HiGPUs} integrates the evolution of the system via the sixth-order Hermite scheme and implements a hierarchy of block timesteps: in particular, the individual timesteps are computed via a combination of the sixth and fourth order Aarseth criterion \citep{Aarseth2003,Nitadori2008};
we set the respective accuracy parameters to $\eta_{\rm sixth}=0.45$,
$\eta_{\rm fourth}=0.01$ \citep[for details, see][]{Capuzzo-Dolcetta2013}.  The minimum and maximum possible values in the hierarchy  are
chosen as $\Delta t_{\rm min}=2^{-29}\approx 1.86\times 10^{-9}$ and $\Delta t_{\rm max}=2^{-6}=0.015625$. {  We set the softening parameter to $\varepsilon=10^{-4}$; such small softening avoids the formation of stellar binaries; at the same time it allows to follow the evolution of the BHB (when present) limiting errors in the energy conservation. Note that such a small softening is required to correctly model the interaction between the BHB and its surrounding stars. It will also slightly reduce the relaxation time of the surrounding stellar distribution, which is a numerical error. However, this error will be small since the relaxation time (see section \ref{sec:rel}) depends linearly on the particle number, $N$, and only logarithmically on the force softening, $\varepsilon$ \citep{Dehnen2011}.}

%
The evolution of the relative energy error of the whole system, i.e.
\begin{equation}\label{eq:energy}
{\left|\frac{\Delta E}{E_{\rm i}}\right|}={\left|\frac{E-E_{\rm i}}{E_{\rm i}}\right	|}
\end{equation}
for runs with $\gamma=1$ and  $e=0.7$ is displayed in Figure~\ref{fig:en}; here  $E_{\rm i}$ is the initial  energy of the whole system, while $E$ is the same quantity evaluated at a given time $t$. The figure shows that energy is well conserved for the entire duration of the simulation when the BHB is not included, as the relative energy error is always below $10^{-6}$; this is true for all simulations without a BHB. 
When a BHB is present, however, the energy error suffers a sudden increase around the time of binary formation ($t\approx t_f$). This is most likely due to the large number of encounters experienced by the binary at this time.  Similar energy errors are obtained in the other simulations, and the energy error reaches values of a few $\times 10^{-3}$ at most.  %

Such energy errors do not invalidate our results regarding the morphology of the remnant since they can be attributed to energetic slingshot ejections of stars which then leave the system. However, they may affect the evolution of the binary parameters. Fig.~\ref{fig:en} also shows the BHB binding energy in run MC7b; here the binary binding energy is at least an order of magnitude larger than the global energy error at any given time, and this holds for all simulations with a BHB.  

\section{Theory}\label{sec:general}

\subsection{BHB evolution}\label{sec:def}

\begin{table}
  \centering
  \caption{Characteristic scales of the binary evolution in the simulations. First column: run identifier; second column: value of $a_f$; third column: time at which the BHB reaches $a_f$; fourth column: value of $a_h$; fifth column: time at which the BHB reaches $a_h$. See the text for further details. }
  \label{tab:a}
  \begin{tabular}{lcccc} 
    \hline
         Run     &  $a_f$ & $t_f$ & $a_h$ & $t_h$  \\
    \hline
LC5b & 0.165  & 342      & 0.0100  & 365 \\
LC7b & 0.152  & 238      & 0.0101  & 259 \\
LC9b & 0.137  & 127      & 0.00841 & 156 \\
MC5b & 0.097  & 368      & 0.00709 & 375 \\
MC7b & 0.096  & 240      & 0.00730 & 246 \\
MC9b & 0.083  & 124      & 0.00673 & 131 \\
HC5b & 0.051  & 408      & 0.00429 & 410 \\
HC7b & 0.045  & 254      & 0.00436 & 256 \\
HC9b & 0.038  & 118      & 0.00419 & 119 \\
    \hline
  \end{tabular}
\end{table}

The evolution of BHBs can be divided into different phases.
Initially, dynamical friction dominates the MBHs evolution and leads to the formation of a bound pair; if the BHB is equal-mass, this roughly happens when the BHB semimajor axis $a_{\rm b}$ drops below $a_f$, i.e. the separation at which the stellar mass $M_\ast$ enclosed in the binary orbit is about twice the mass of one MBH \citep{BinneyTremaine2008}:
\begin{equation}
M_\ast(a_f)=2M_\bullet.
\end{equation}
Around this time,  the merger process can be generally assumed to be complete; in what follows we will use $t_f=t(a_f)$ as a reference time for the end of the merger
\footnote{
The same $t_f$ is used for simulations with the same initial orbit and density profile but with only one or no MBHs, as  the merger timescale is approximately independent of the presence of the massive bodies.
}.

Starting from $t_f$, dynamical friction coupled with slingshot ejections of stars rapidly shrinks the binary and empties the BHB loss cone for the first time; in addition, a core is carved in the stellar distribution as a result of slingshot ejections. When $a_{\rm b}$ reaches 
\begin{equation}
a_h=\frac{G\mbin}{8\sigma^2}, 
\end{equation}
(where $\mbin=2M_\bullet$ is the BHB mass, and $\sigma$ is the one-dimensional velocity dispersion of the field stars) the binary is said to be `hard' as its binding energy per unit mass exceeds the mean stellar binding energy per unit mass; from this moment, stars ejected by the BHB are able to escape the galactic potential. Around  time $t_h=t(a_h)$ the BHB shrinking considerably slows down as the loss cone has been emptied and any further hardening  depends on the loss cone repopulation rate. We define the time dependent BHB hardening rate as
\begin{equation}
s(t)=\frac{\rm d}{{\rm d }t}\left(\frac{1}{a_{\rm b}}\right);
\end{equation}
since the BHB mass is constant in time, $s$ is an estimate of the BHB energy loss. 

The hardening process continues  until emission of GWs becomes effective and leads the BHB to its final coalescence. 
The significant scales ($a_f,a_h$) in the BHB evolution and their associated times are listed in table \ref{tab:a}.

\subsection{Two-body relaxation}\label{sec:rel}
\begin{table*}
  \centering
  \caption{Relaxation timescales in N-body units at different radii. The first column shows the name of the run, while the next columns list the relaxation timescales at the radius enclosing a different fraction $m_{\rm encl}$ of the total stellar mass, i.e. (from left to right) the 0.5\%, 10\%, 25\%, 50\%, 75\%; the relaxation timescales are computed at $t\approx t_h$ for runs including the BHB and at the corresponding time for runs without MBHs.}
  \label{tab:rel}
  \begin{tabular}{lccccc} 
    \hline
Run  &$m_{\rm encl}=0.5\%$&$m_{\rm encl}= 10\%$&$m_{\rm encl}= 25\%$&$m_{\rm encl}= 50\%$&$m_{\rm encl}= 75\%$\\
    \hline
LC7  & 600    & 6,900 & 28,500 & 260,000 & 2.5M\\ 
LC7b & 1,900  & 7,500 & 29,000 & 280,000 & 2.3M\\
MC7  & 160    & 4,300 & 19,000 & 220,000 & 2.0M  \\
MC7b & 850    & 4,500 & 20,500 & 191,000 & 2.0M  \\
HC7  & 50     & 2,000 & 11,000 & 143,000 & 2.3M\\
HC7b & 370    & 2,200 & 13,000 & 142,000 & 2.1M\\
    \hline
  \end{tabular}
\end{table*}

Two-body relaxation operates on a timescale  $T_{\rm rel}$ that strongly correlates with the number of particles in the system, roughly as $T_{\rm rel}\propto N/\log N$, and it is known to exceed the Hubble time in almost all sufficiently luminous galaxies. However the limited number of particles ($N=512k$) in our simulations results in a significantly smaller relaxation time. In table~\ref{tab:rel} we list the relaxation timescale of systems with $e=0.7$, both with and without the BHB, at different shells of enclosed mass when $a_{\rm b}\approx a_h$; the relaxation time is computed from simulation snapshots as
\begin{equation}\label{eq:trel}
T_{\rm rel}=\frac{0.34 \sigma^3}{G^2m_\ast\rho\ln \Lambda}
\end{equation} 
\citep{Spitzer1987}, where $\sigma$ is the one-dimensional velocity dispersion, $m_\ast$ is the stellar mass, $\rho$ is the averaged density within the shell and $\ln \Lambda$ is the Coulomb logarithm, computed as $\ln \Lambda=\ln( \mbin /m_\ast)$ within the BHB influence radius (if the BHB is present) and as $\ln \Lambda=\ln(r_{80}/\varepsilon)$ otherwise; $r_{80}$ is the radius enclosing 80\% of the stellar mass. 

In our runs, the relaxation time is significantly longer than the simulation time at radii containing a fraction of the total stellar mass of the order of 50\% or above, but this no longer holds at smaller radii, in particular when the BHB is not present and the progenitor galaxies are more concentrated\footnote{
The relaxation time depends primarily on the density and the velocity dispersion of the system. Within the half mass radius, the density at a given fraction of enclosed mass significantly increases with $\gamma$, while $\sigma^3$ exhibits  a weaker growth; as a result, at small scales, the relaxation time becomes shorter if the galaxy concentration is enhanced. However, at radii enclosing more than half of the total mass, the density increase with $\gamma$ becomes milder, and can be hindered by the comparable growth of $\sigma^3$. For this, at large scales, the relaxation time may be found to grow with increasing $\gamma$ (Table \ref{tab:rel}).
}. The  computation of $T_{\rm rel}$ will enable us to disentangle the effects of spurious two-body relaxation from the consequences of the merger and BHB evolution.

\subsection{Computation of triaxiality}

The shape of the galactic merger remnant can be determined by computing the ellipsoid that best approximates the stellar distribution at a given distance. If $a>b>c$ are the axes of this ellipsoid, a deviation from perfect sphericity can be evaluated quantitatively as the departure of  $b/a$, $c/a$ from unity.  When a single MBH or a BHB is present, we evaluated the axes of the ellipsoid  using all the stars enclosed within a sphere of radius $r$ centered on the MBH or on the BHB centre of mass; if the BHB is not present, the centre of the system is instead assumed to be the centre of mass of the 75\% innermost particles; we verified that the different evaluations of the spheroid's centre  do not affect the computation of the axis ratios. The procedure we adopt for the evaluation of the remnant shape is the same as in \citet{Katz1991} and \citet{Antonini2009}.

The first order axis ratios  are determined from the eigenvalues ($\xi, \eta, \theta$) of the inertia tensor $I$: $\xi=\sqrt{I_{11}/I_{\rm max}}$, $\eta=\sqrt{I_{22}/I_{\rm max}}$, $\theta=\sqrt{I_{33}/I_{\rm max}}$, where $I_{\rm max}=\max(I_{11},I_{22},I_{33})$. In order to get a better accuracy, we iterate the procedure and we computed new axis ratios by considering only particles enclosed in an ellipsoidal volume with the previously computed ($\xi, \eta, \theta$), i.e. all particles $\rm i$ located in $(x_{\rm i}, y_{\rm i}, z_{\rm i}, )$ satisfying
\begin{equation}
q_{\rm i}^2=\left(\frac{x_{\rm i}}{\xi}\right)^2+\left(\frac{y_{\rm i}}{\eta}\right)^2+\left(\frac{z_{\rm i}}{\theta}\right)^2<r^2.
\end{equation} 
The procedure is iterated until an accuracy of $10^{-5}$ is achieved in the computation of the axis ratios; the ellipsoid is free to rotate about its centre at each iteration. Finally, we define the ellipsoidal axes $a>b>c$ such that ($1, b/a, c/a$) are equal to ($\xi, \eta, \theta$) in the right order. 
It is then possible to compute the triaxiality parameter $T$ of the system within $r$ as:
\begin{equation}
T=\frac{a^2-b^2}{a^2-c^2};
\end{equation}
$T$ is a quantity used to describe the deviation of a system from perfect sphericity and it can vary in the range $[0, 1]$:
\begin{enumerate}
\item if $0\leq T <0.5$ the spheroid is oblate;
\item if $T=0.5$ the system is said to have maximum triaxiality;
\item if $0.5 < T \leq 1$ the spheroid is prolate.
\end{enumerate}
In the description of the results, we will make use of both the axis ratios ($b/a$, $c/a$) and the  triaxiality parameter $T$ to characterize the  shape of the remnant. {  We stress that the shortest axis of the spheroid  typically lies perpendicularly with respect to the merger plane (when a BHB is present, the merger plane is always parallel to the BHB orbital plane).}

\section{Results}\label{sec:res}

\begin{figure*}
\includegraphics[trim={0cm 0cm 0cm 0cm},width=\textwidth]{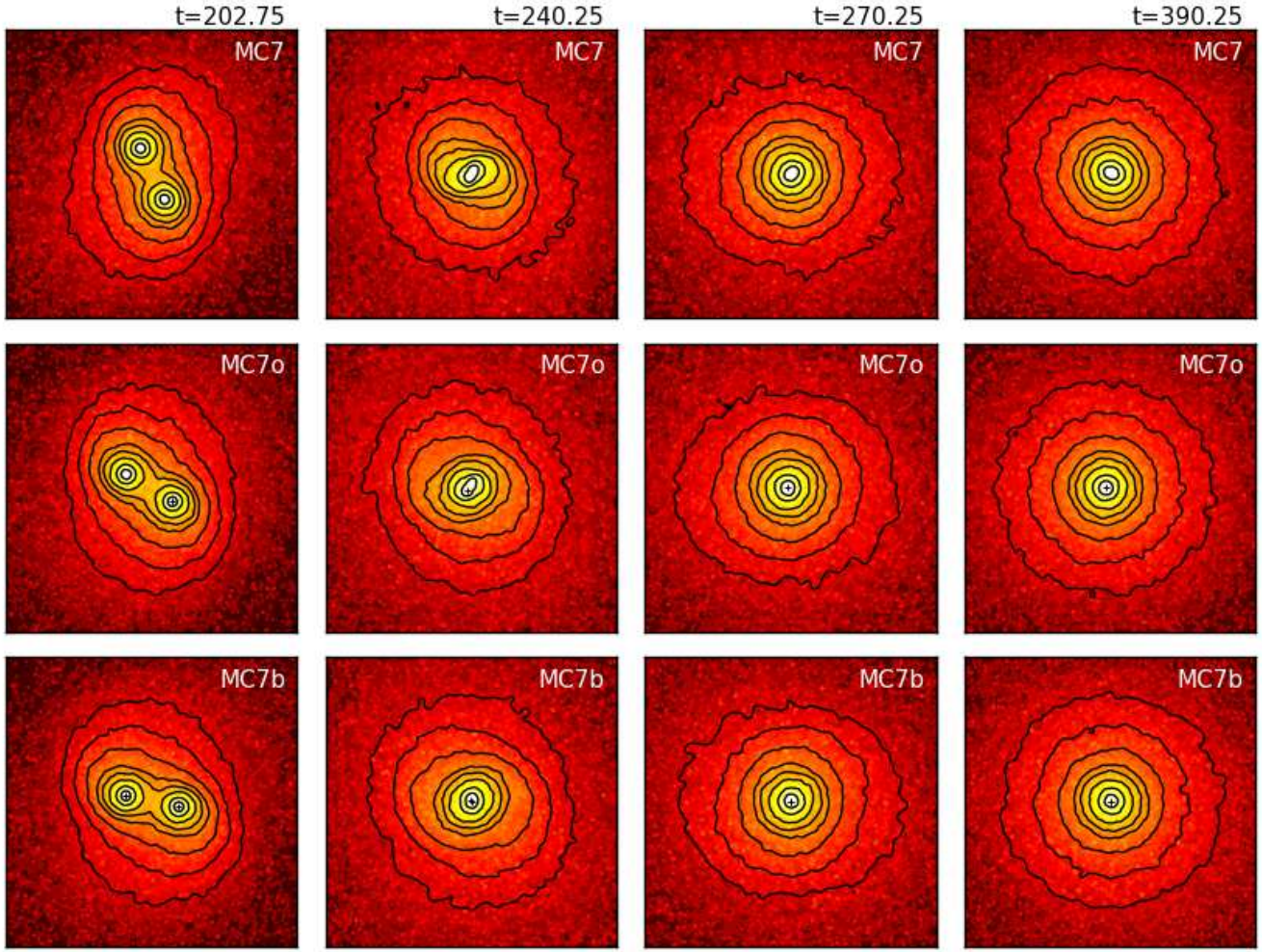} 
 \caption{ The snapshots show the galaxy merger evolution in three different runs: MC7 (no MBHs), MC7o (one MBH) and MC7b (two MBHs); small black crosses mark the position of the MBHs and {  the colour code refers to the projected mass density, ranging from $\approx500$ to $\approx 2.5\times10^5$ particles per squared $N$-body unit}. The $N-$body time  associated with each snapshot is shown at the top of the image; each box is 13 $N$-body units wide.}
    \label{fig:splash}
\end{figure*}

\begin{figure}
\includegraphics[angle=270,trim={9.5cm 0cm 1.2cm 0cm},width=\columnwidth]{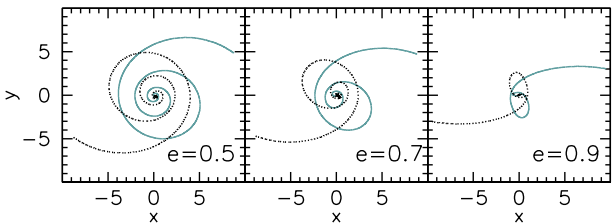} 
 \caption{Trajectories of the two MBHs in the merger plane for runs MC5b, MC7b, MC9b, i.e. with  $e=0.5,0.7,0.9$, from left to right. We only show the orbits for simulations with $\gamma=1$, as the large-scale  MBH paths  only weakly depend on the density of the progenitor galaxies. In this Figure and in the following, distances are in scalable $N-$body units.}
    \label{fig:path}
\end{figure}

\begin{figure}
\center
\includegraphics[trim={1cm 0cm 7cm 12cm},width=.78\columnwidth]{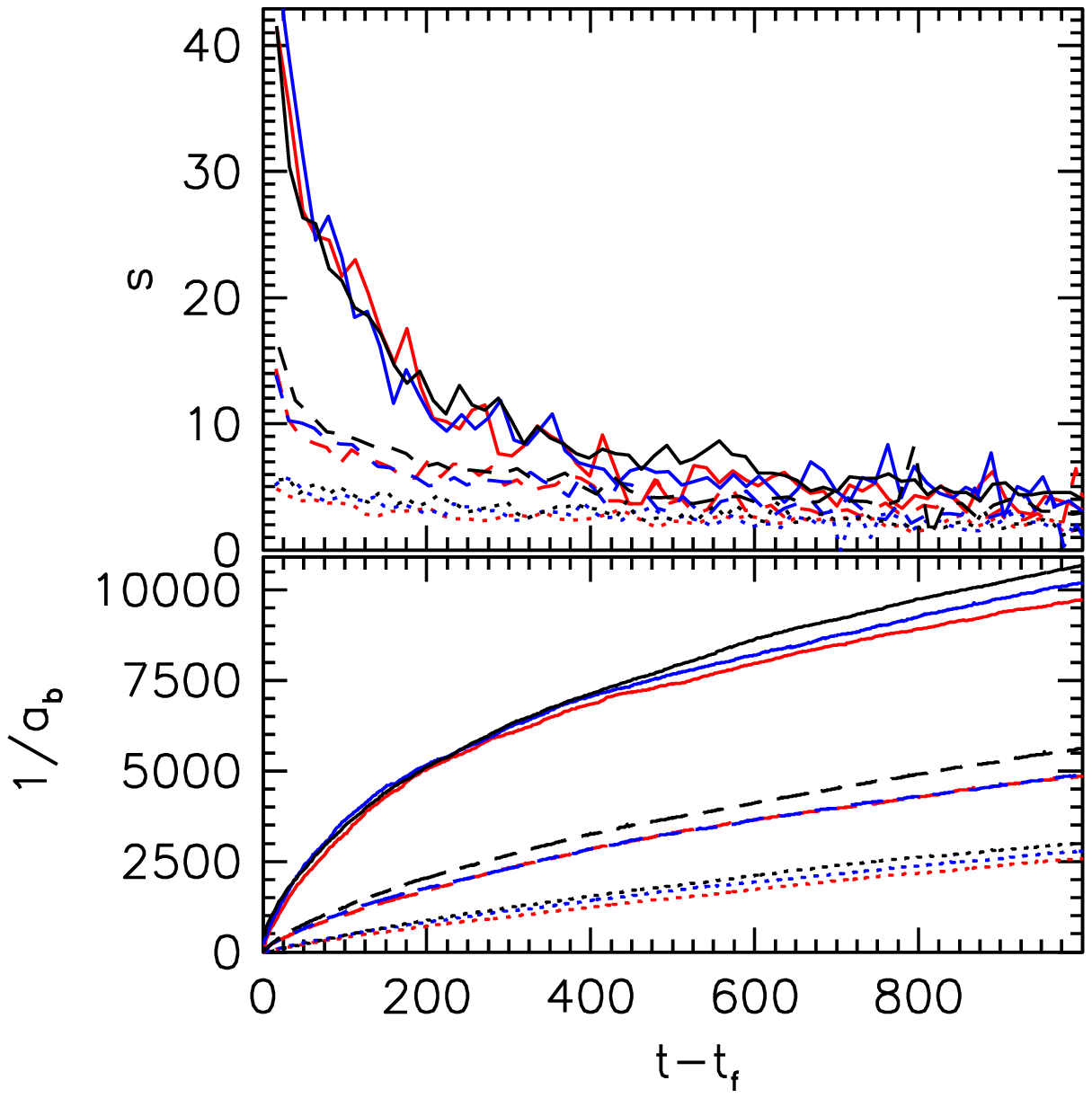}
\caption{Time evolution of the BHB hardening rate ($s$, top panel) and the  inverse semimajor axis of the binary ($a_{\rm b}^{-1}$, bottom panel). Dotted lines: $\gamma = 0.5$; dashed lines: $\gamma=1$; solid lines: $\gamma =1.5$. Red lines: $e=0.5$; blue lines: $e=0.7$; black lines: $e=0.9$.  Time on the horizontal axis is given starting from $t_f$. }  \label{fig:s}
\end{figure}

\subsection{BHB evolution}
{ Figure~\ref{fig:splash} shows the evolution of the galactic collision in the merger plane for runs MC7, MC7o and  MC7b, i.e. three simulations with the same merger orbit and galaxy density profile, but including respectively zero, one and two MBHs. 

The trajectories of the two MBHs in the plane of the merger for different values of the merger eccentricity are instead shown in Figure~\ref{fig:path}; we stress that } when the MBHs are not included, the orbital evolution of the centres of density  are not significantly different to the analogous MBH paths in Figure~\ref{fig:path}.
Table~\ref{tab:a} shows that the merger is faster if the orbital eccentricity is higher, as galaxies experience a closer pericentre passage.

Figure~\ref{fig:s} displays the temporal evolution of the BHB hardening rate and the inverse of the semimajor axis. The BHB hardening rate does not show any clear relation with the initial orbital eccentricity (Figure~\ref{fig:s}, top panel) { but the binary appears to shrink slightly faster when the merger eccentricity is higher (Figure~\ref{fig:s}, bottom panel).}

The hardening rate  strongly depends on the density slope $\gamma$ of the progenitors: if the system is more compact, more stars are initially available for three-body interactions with the BHB in the inner region of the remnant and the BHB hardening is more efficient, as already discussed in \citet{Sesana2015} and \citet{Vasiliev2015}. 
 In addition, we note that the BHB hardening rate decreases in time, especially when $\gamma$ is high; {  in particular, $s$ in all our simulations tends to the same value ($s\approx 3 - 4$) towards the end of the runs.} {The decline in the BHB hardening rate has also been observed in a series of recent papers (e.g. \citealt{Vasiliev2015,Sesana2015,Gualandris2017}); according to \citet{Vasiliev2015}  } it is due to the fact that  low-energy orbits are more populated in steeper models, thus more stars are initially found in the BHB loss cone; once most of the low-energy orbits have been depleted, less stars on higher energy orbits are available to interact with the BHB. { We mention that a  slowing down of the hardening rate  might also be caused by a loss of triaxiality in the system, as suggested in \citet{Vasiliev2015}}.

Finally, we note that the trend in the BHB hardening rate we find here is comparable to what found in previous similar studies that refute the final parsec problem \citep[e.g.][]{Khan2011,Preto2011,Gualandris2012}, thus our results confirm that  BHBs should be able to coalesce in less than a Hubble time in most galaxies.

\subsection{Triaxiality of the system}

\begin{figure*}
\includegraphics[trim={1cm 0.3cm 8.75cm 13cm},width=.3\textwidth]{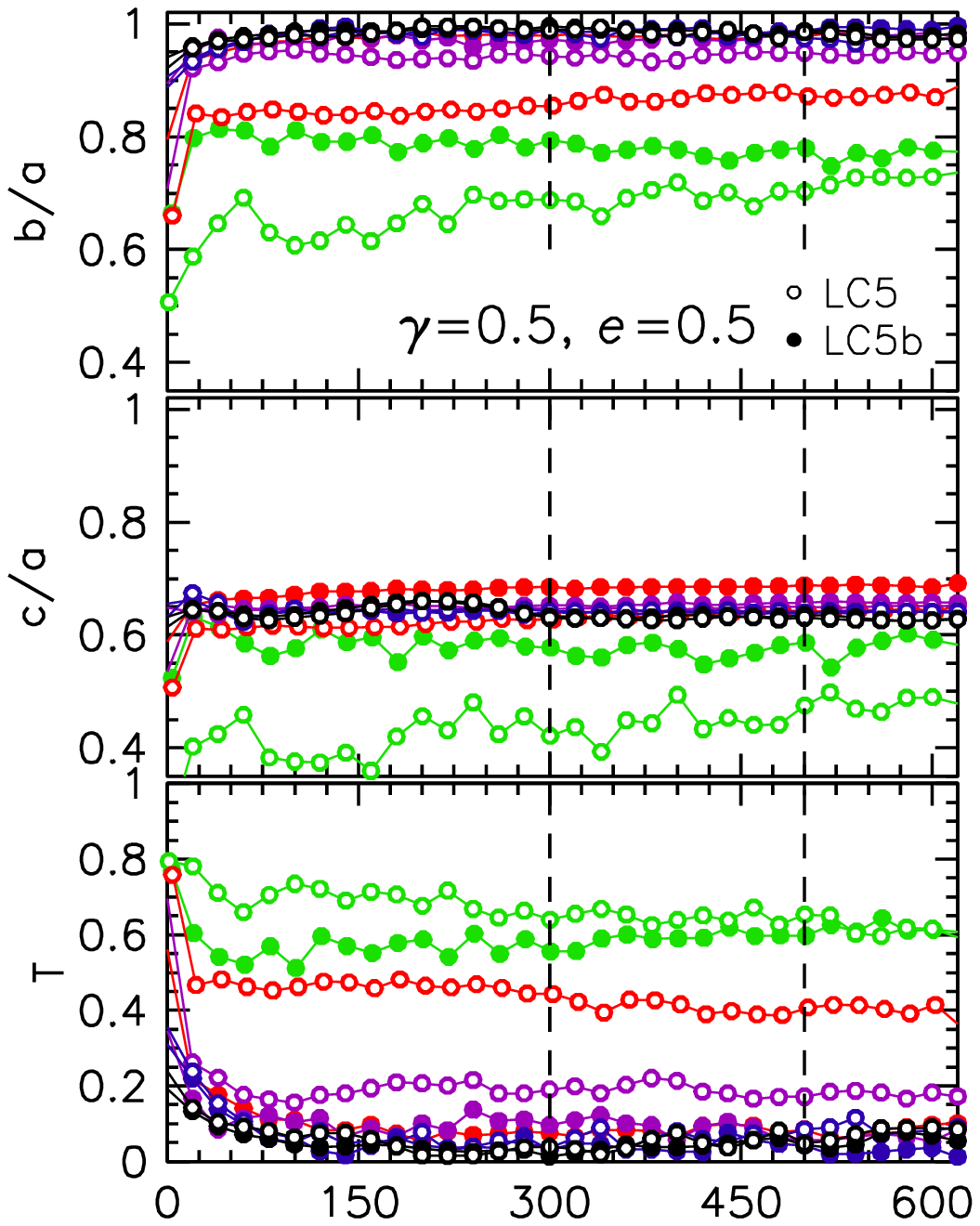} 
\includegraphics[trim={1cm 0.3cm 8.75cm 13cm},width=.3\textwidth]{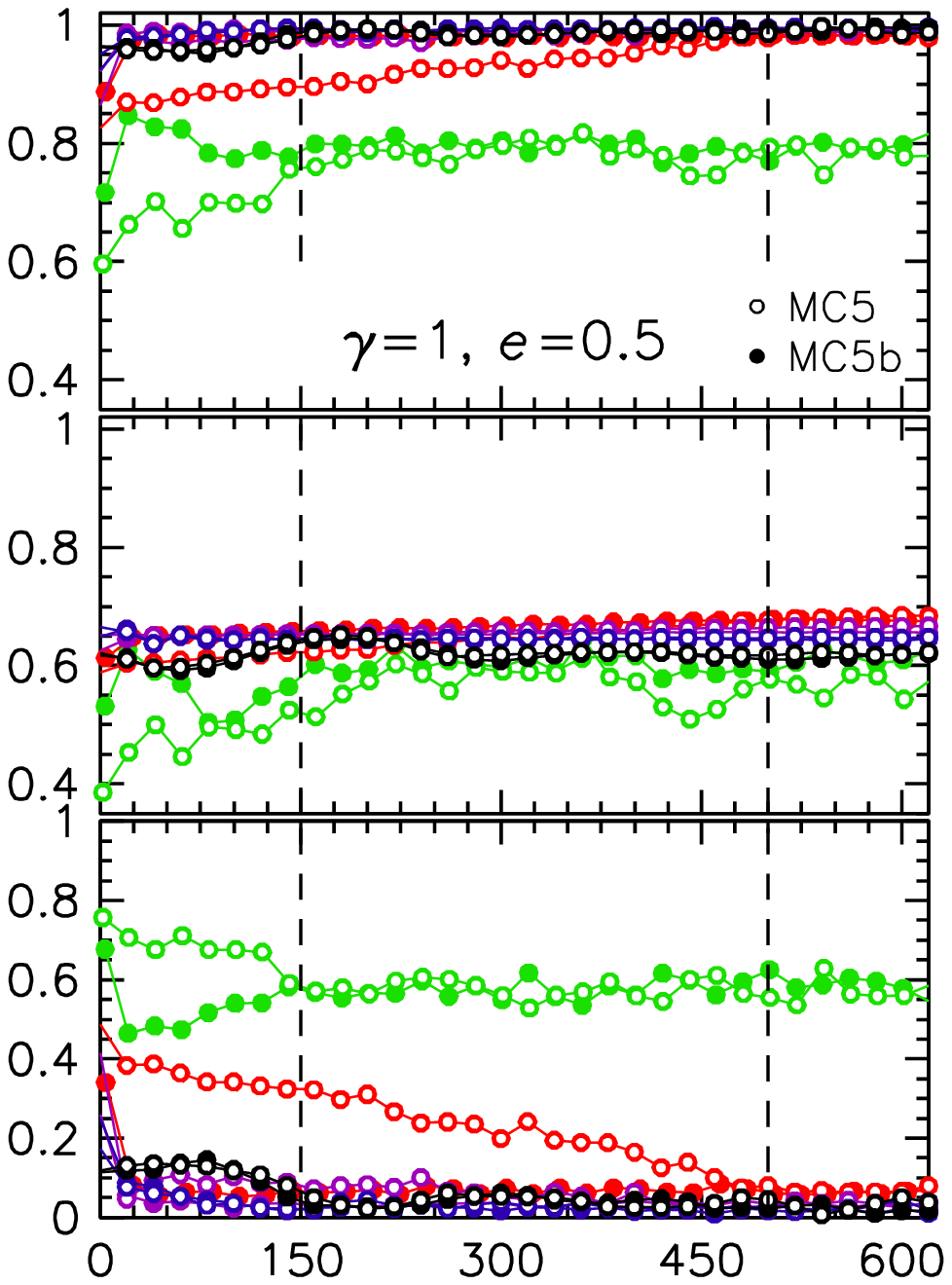}
\includegraphics[trim={1cm 0.3cm 8.75cm 13cm},width=.3\textwidth]{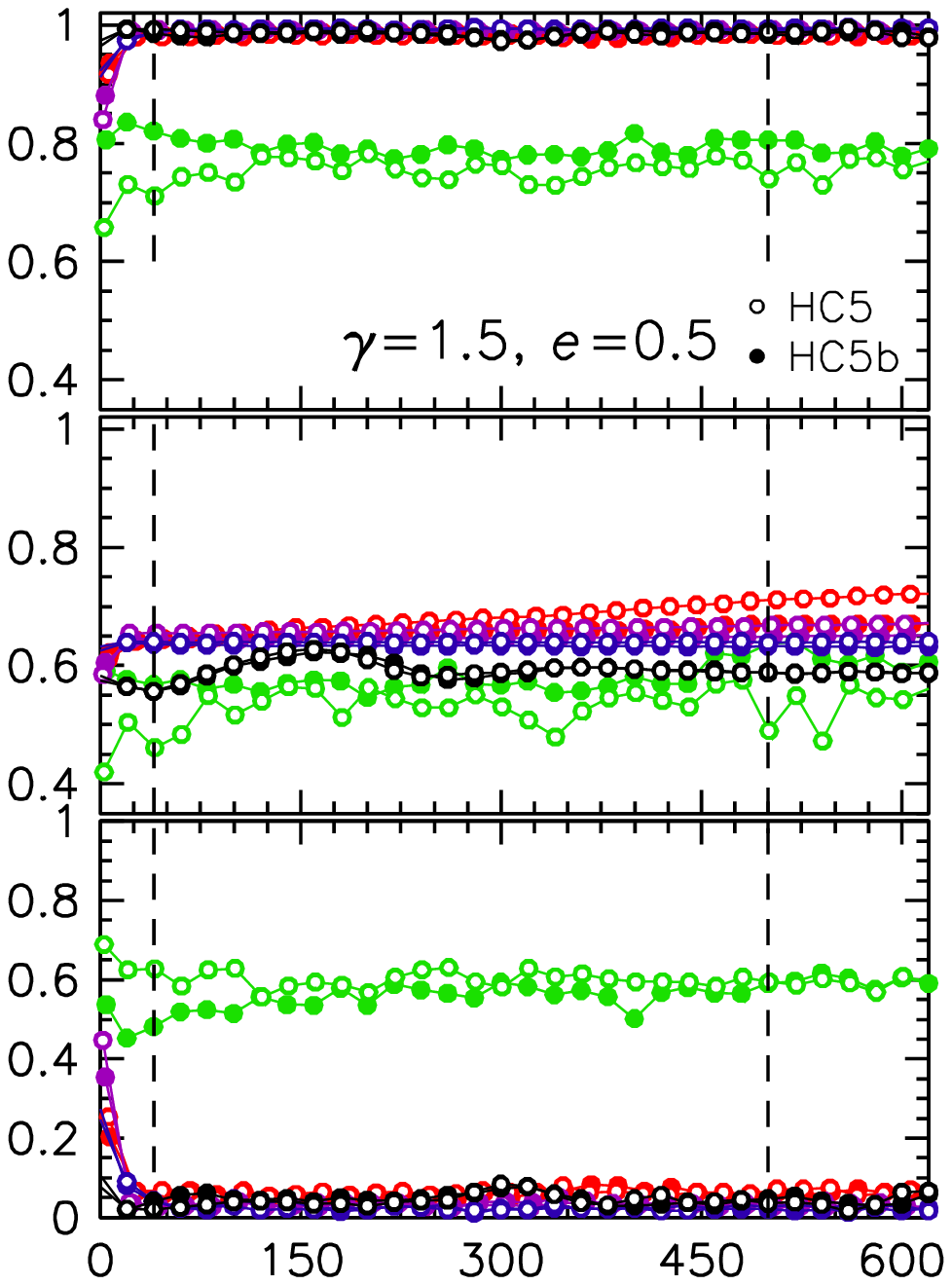}

\includegraphics[trim={1cm 0.3cm 8.75cm 13cm},width=.3\textwidth]{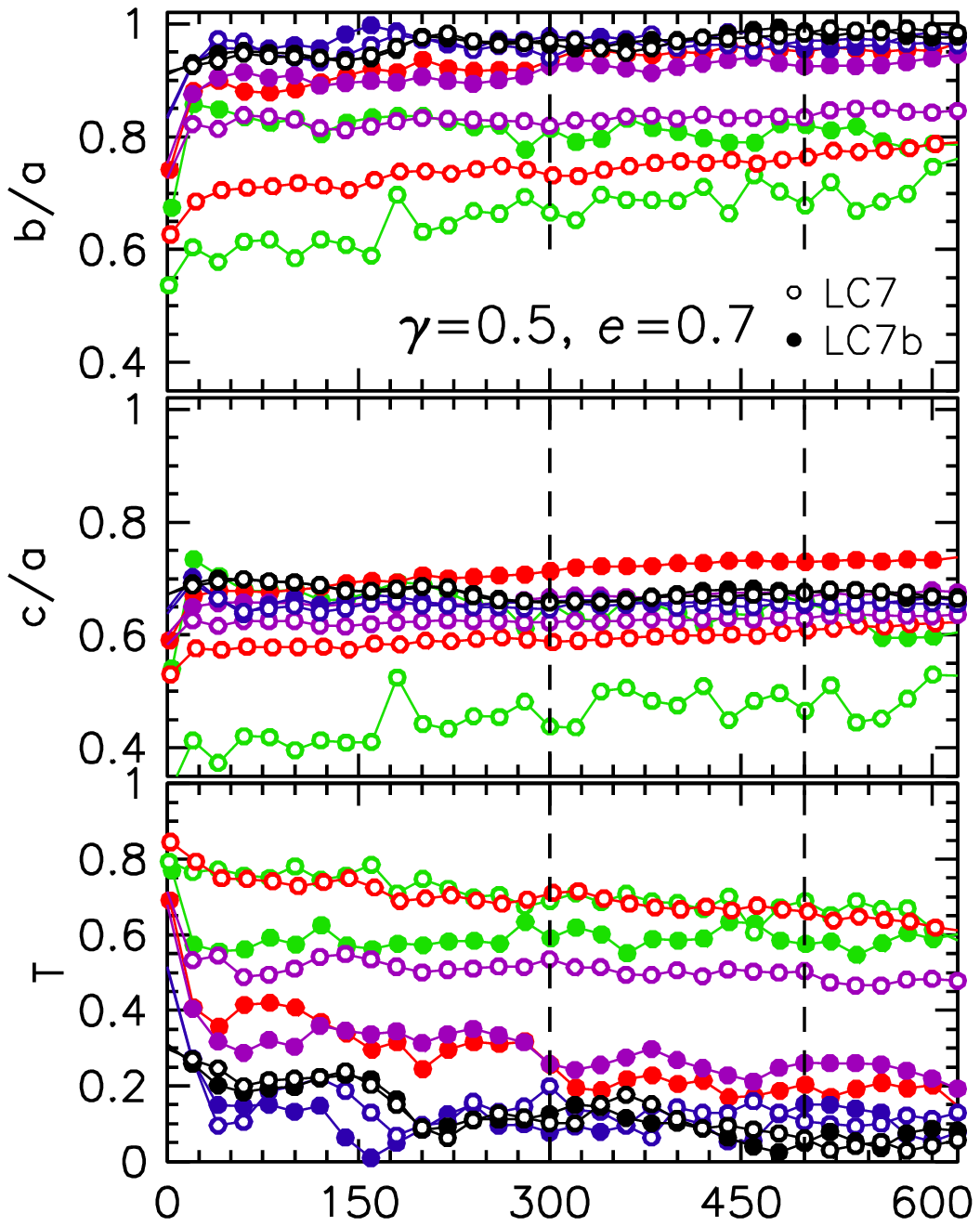} 
\includegraphics[trim={1cm 0.3cm 8.75cm 13cm},width=.3\textwidth]{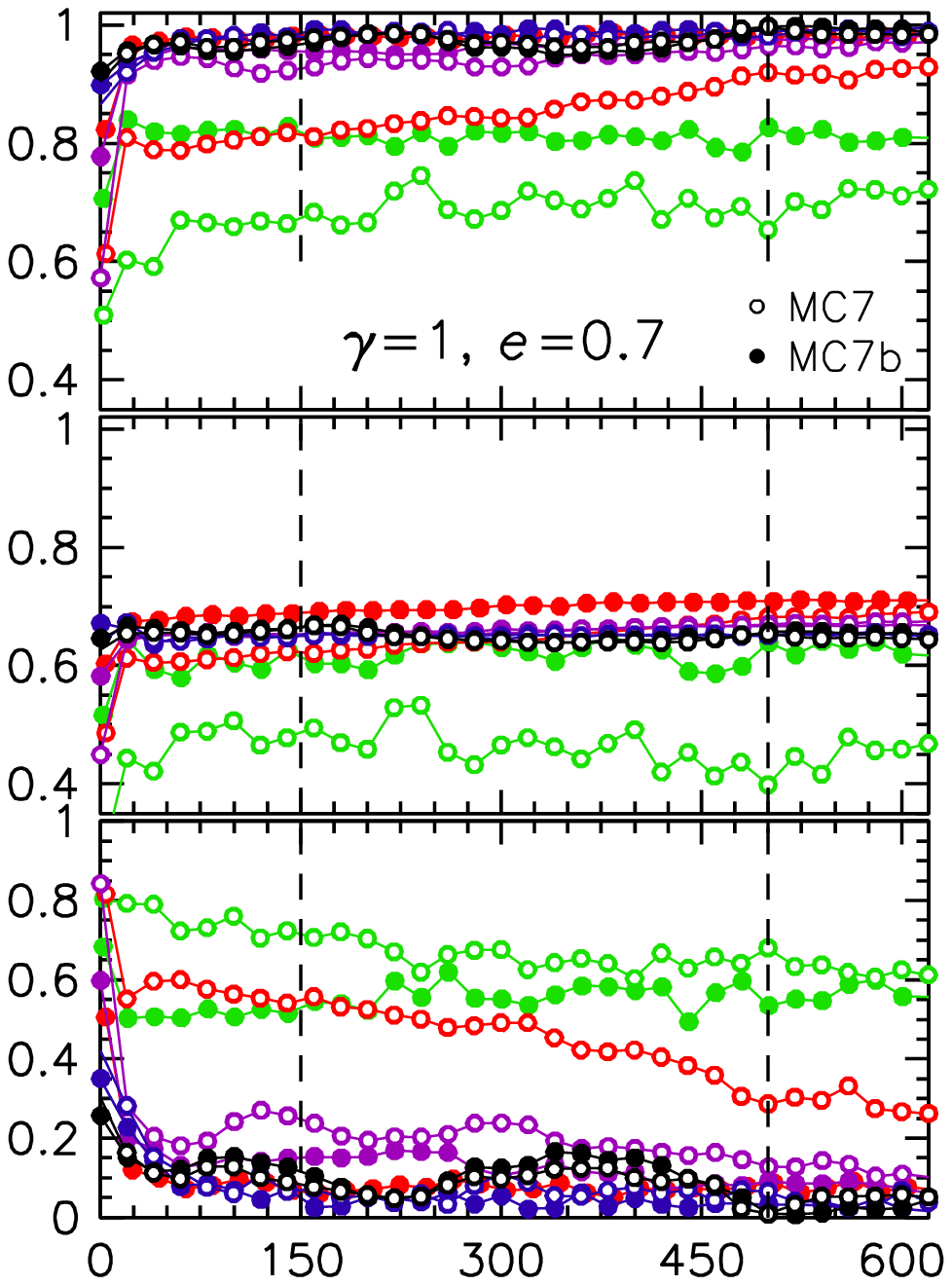}
\includegraphics[trim={1cm 0.3cm 8.75cm 13cm},width=.3\textwidth]{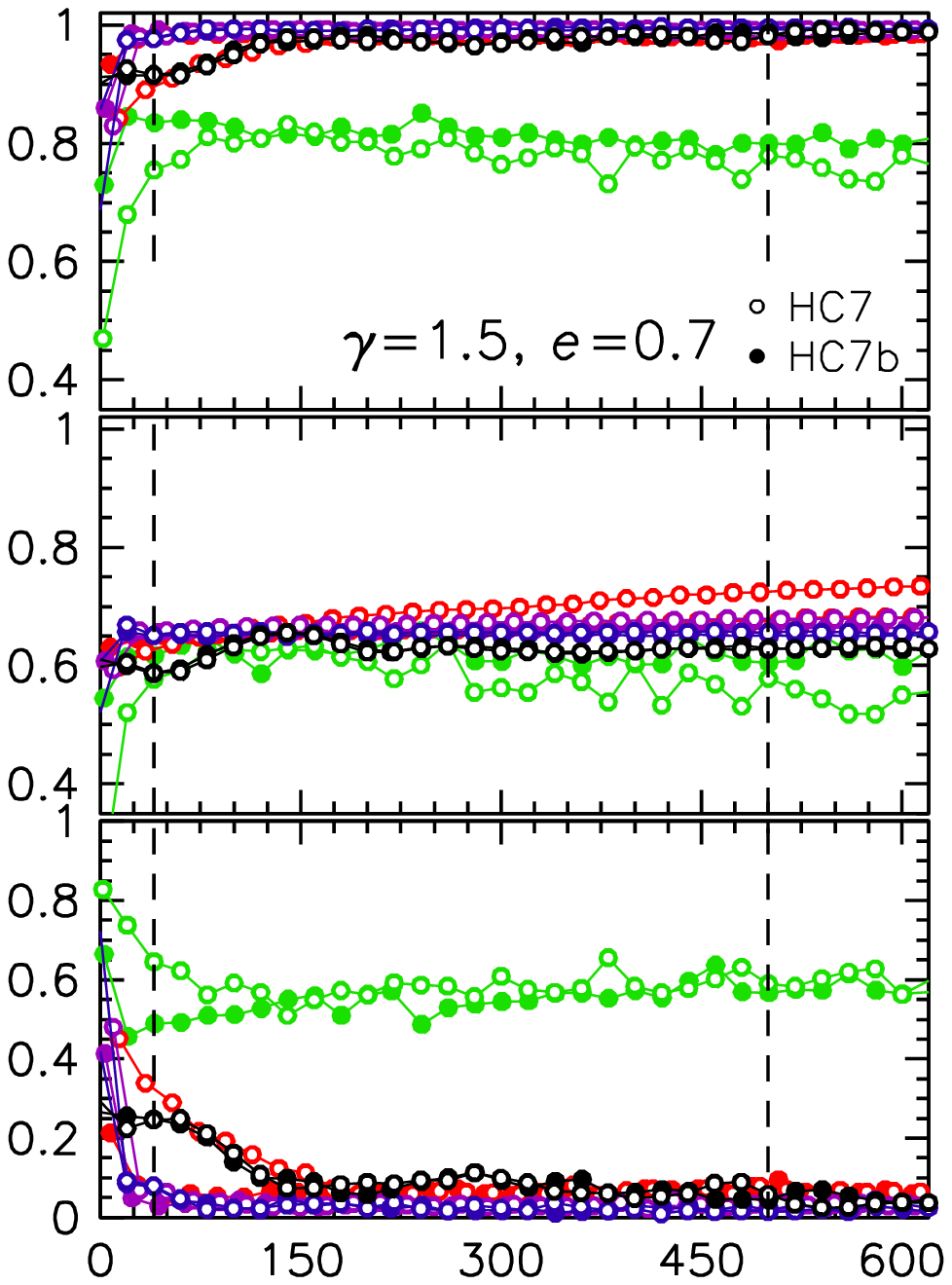}

\includegraphics[trim={1cm 0.3cm 8.75cm 13cm},width=.3\textwidth]{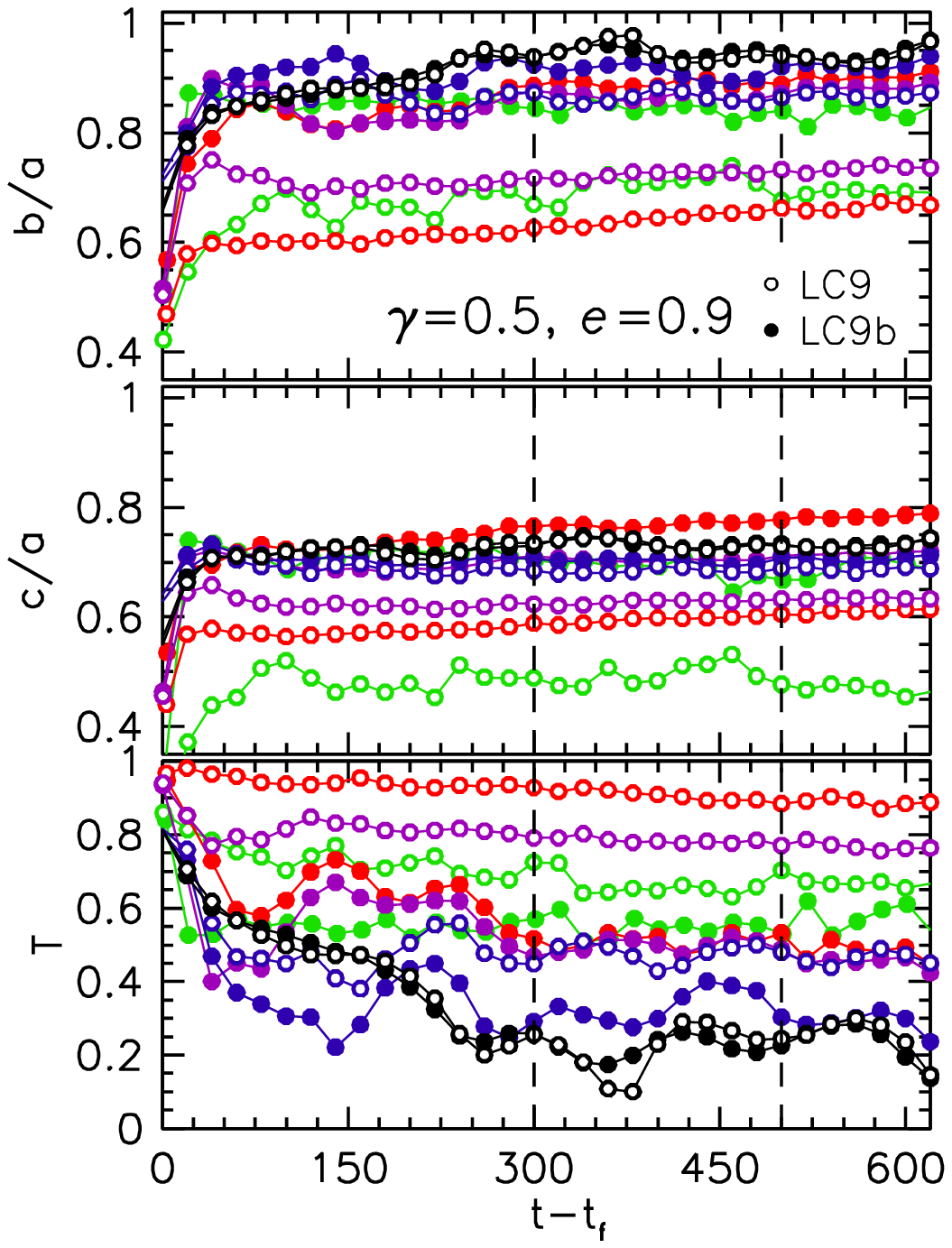} 
\includegraphics[trim={1cm 0.3cm 8.75cm 13cm},width=.3\textwidth]{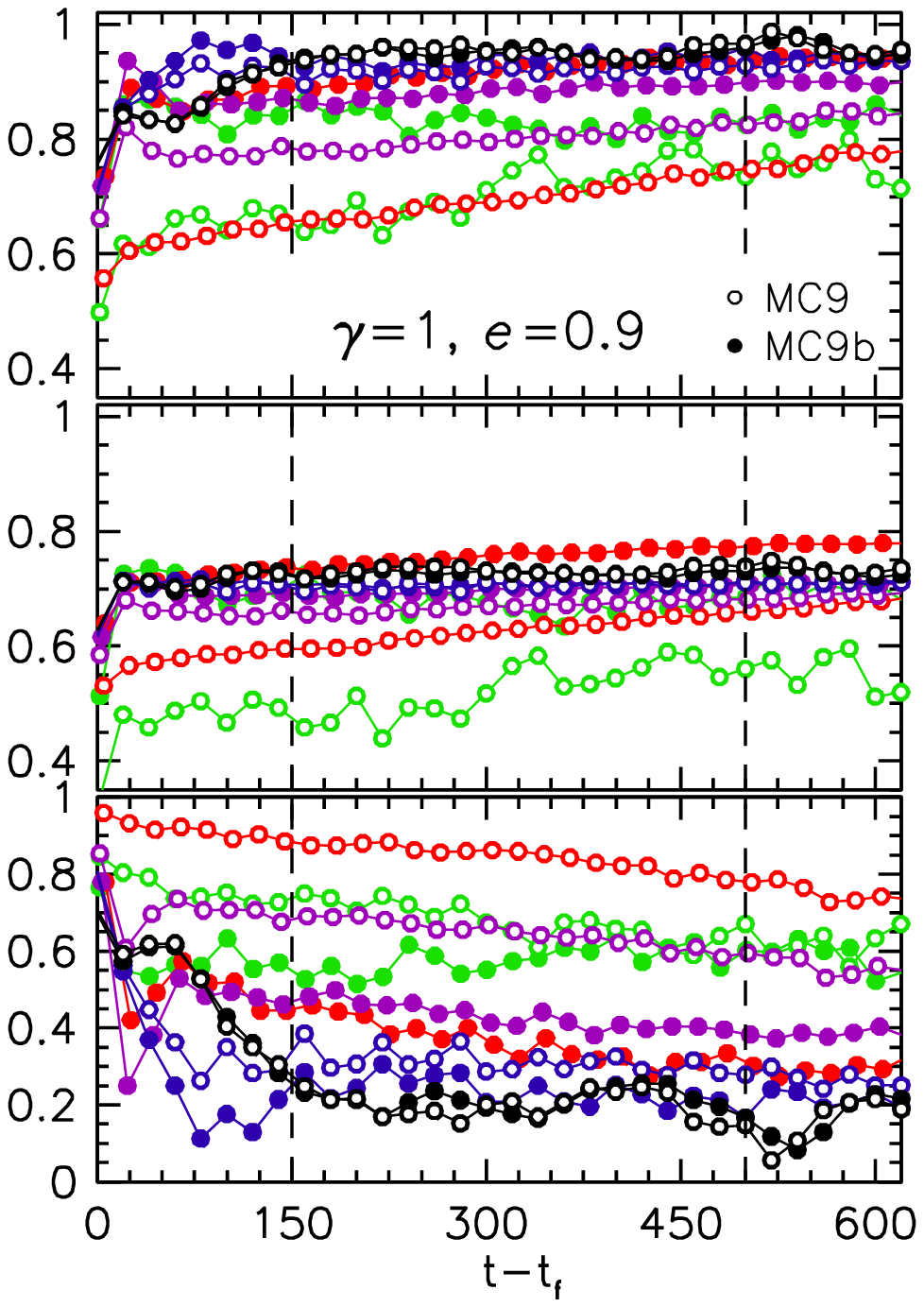}
\includegraphics[trim={1cm 0.3cm 8.75cm 13cm},width=.3\textwidth]{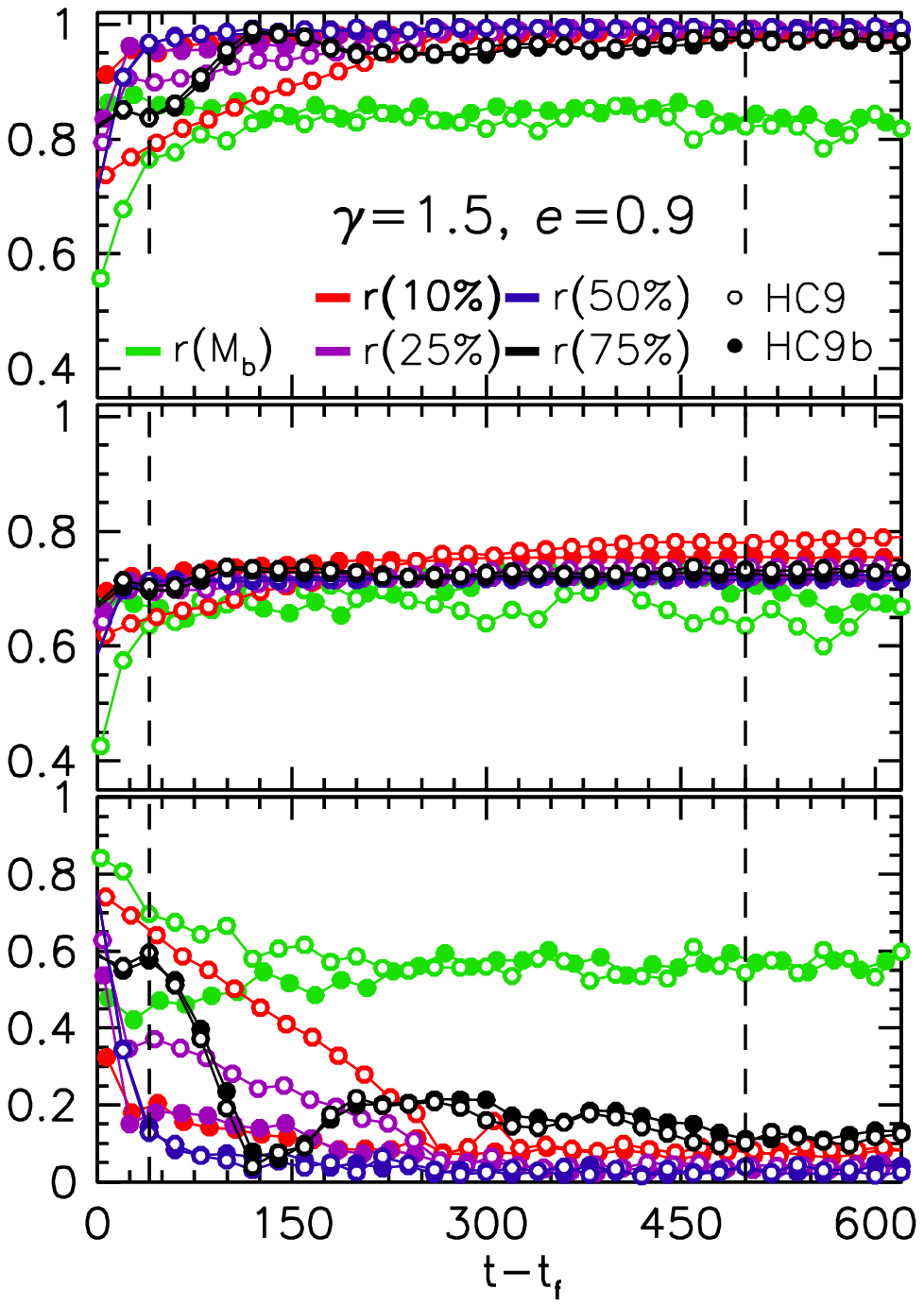}

\vspace*{-2mm}

\caption{ Triaxiality parameter $T$ and axis ratios as a function of time for models with $e=0.5$ (top row), $e=0.7$ (central row) and $e=0.9$ (bottom row); simulations with initial galaxies having  $\gamma=0.5, 1, 1.5$ are shown respectively on the left, central and right-hand column. Each plot consists of three panels showing the temporal evolution of (from top to bottom)  $b/a$, $c/a$ and $T$; these quantities have been averaged over small time intervals to reduce noise; the time evolution is shown starting from $t_f$, i.e. when the galaxy merger is completed. Different lines indicate the parameters computed using particles within a sphere enclosing a fraction equal to the 0.5\%  (green),  10\% (red), 25\% (violet),  50\% (blue) and 75\% (black) of the total stellar mass; simulations including the BHB are shown with filled points, while simulations without MBHs are shown in empty  points. In all plots, the vertical dashed line on the left marks the reference time at which the axis ratios and triaxiality of the structure enclosing the 25\% of the stellar mass is evaluated; the  line on the right shows the reference time for evaluating the morphology of the system at larger scales. Note that the triaxiality increases for increasing orbital eccentricity (top to bottom) and for decreasing concentration (right to left).}
\label{fig:triax_t} 
\end{figure*}

\begin{figure}
\includegraphics[trim={.0cm 0cm 4.5cm 15.5cm},width=.9\columnwidth]{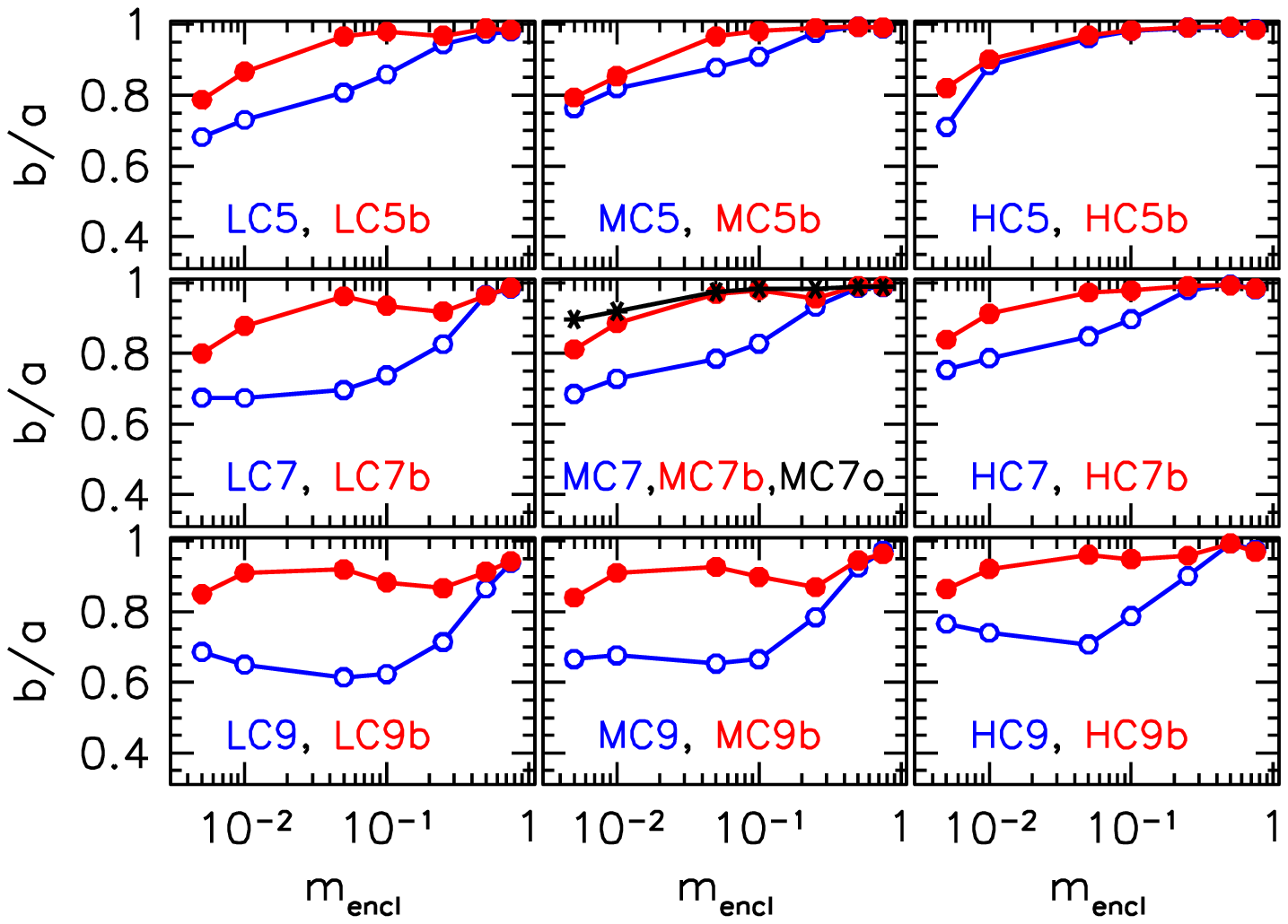}
\includegraphics[trim={.0cm 0cm 4.5cm 15.5cm},width=.9\columnwidth]{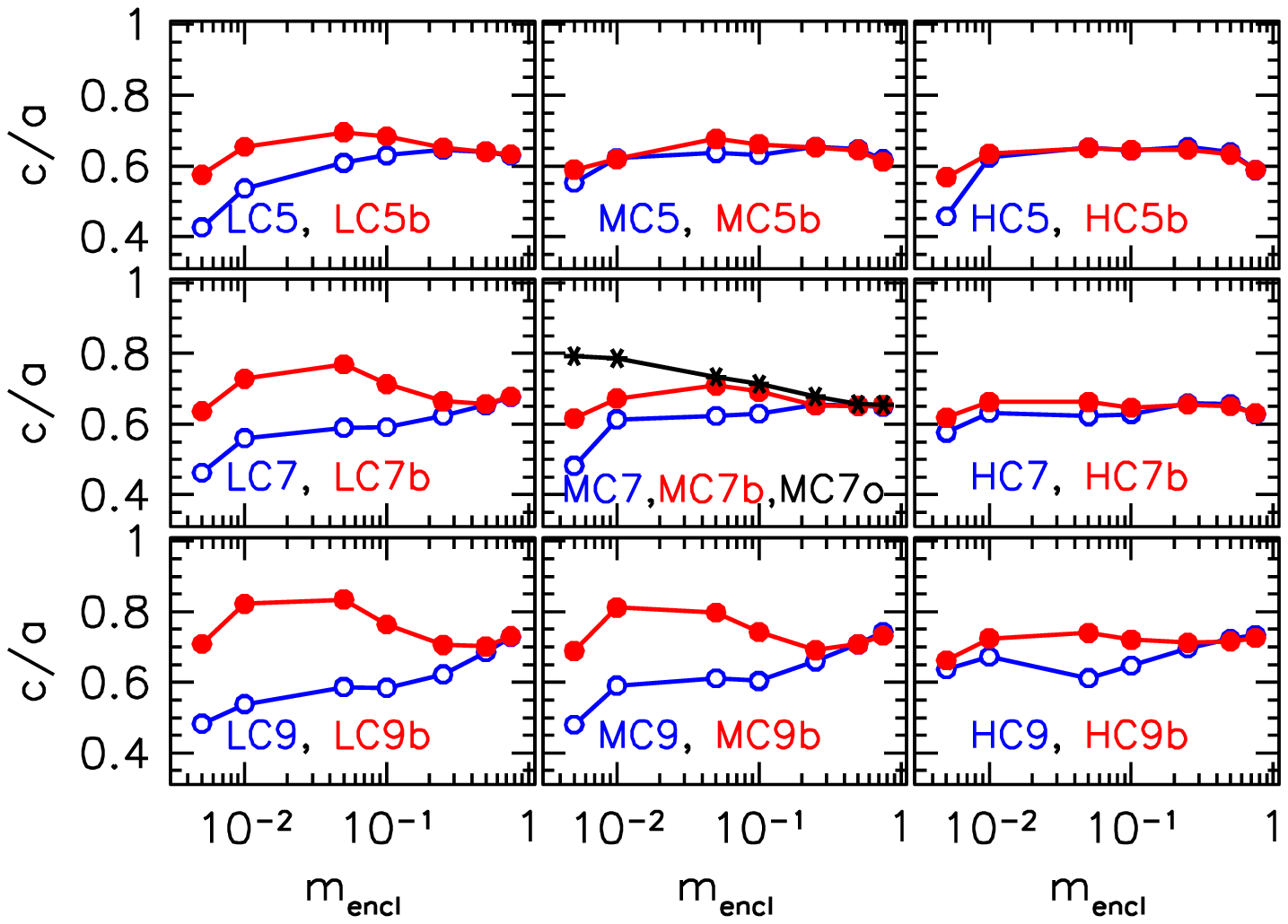} 
\includegraphics[trim={.0cm 0cm 4.5cm 15.5cm},width=.9\columnwidth]{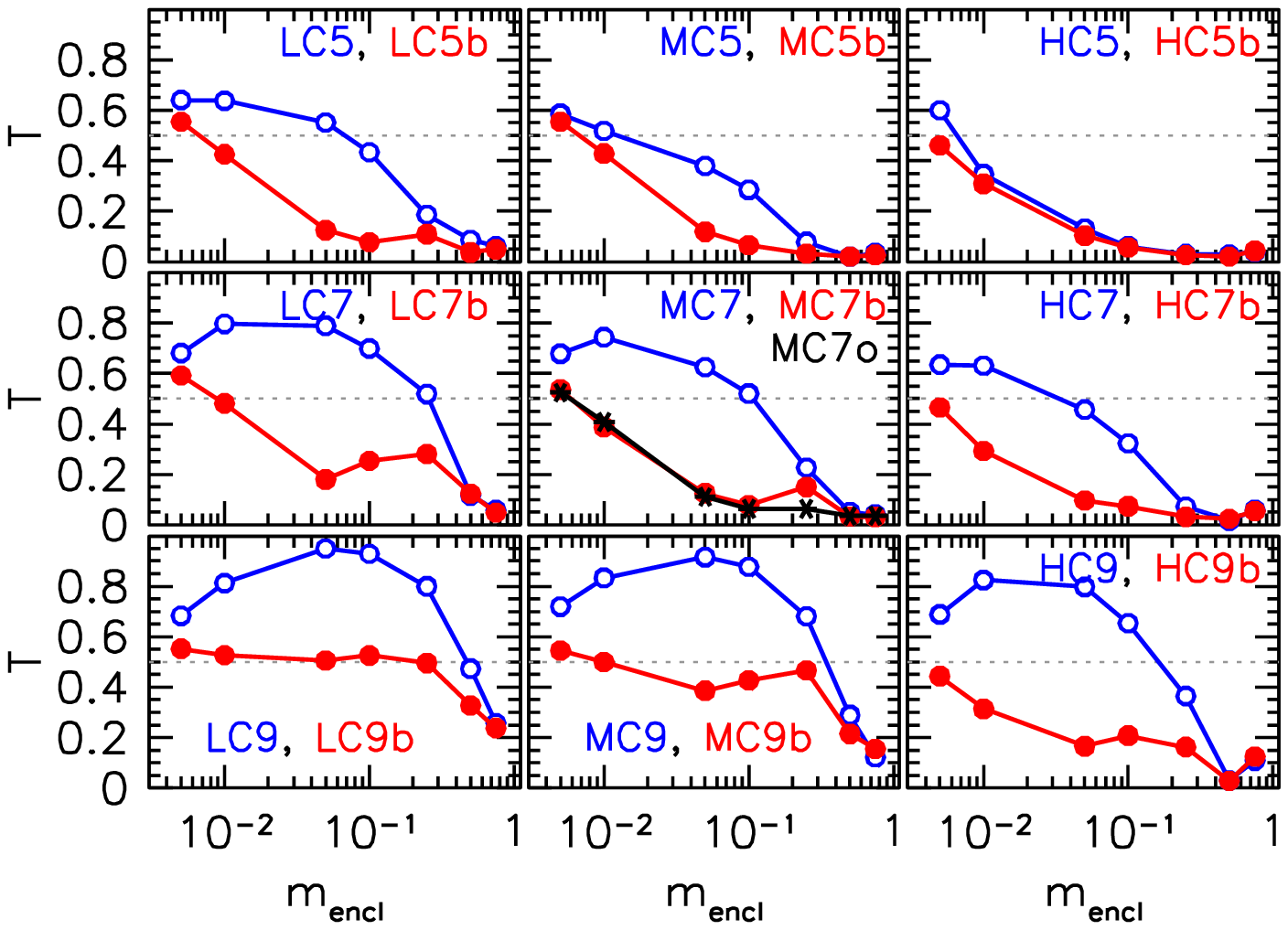} 

\caption{Triaxiality parameter and axis ratios  as a function of the enclosed mass $m_{\rm encl}$; from top to bottom, the plots show $b/a$, $c/a$ and $T$  for runs including the BHB (filled red points), omitting it (blank blue  points) { and for the run with only one MBH (run MC7o, black asterisks in the central panels)}. Each panel is labelled with the name of the runs shown: {  from left to right, panels show runs with increasing concentration ($\gamma = 0.5, 1$ and 1.5), while from top to bottom, panels show runs with increasing eccentricity ($e=0.5, 0,7$ and 0.9) as in Fig. \ref{fig:triax_t}.}}
\label{fig:triax_m} 
\end{figure}

In this section we describe the morphology of the merger relic and its dependence on  (i) the presence of the BHB and (ii) the initial conditions of the merger. 
Figure~\ref{fig:triax_t} shows the temporal evolution of the axis ratios and triaxiality parameter $T$, while Figure~\ref{fig:triax_m} shows how the same quantities vary as a function of the enclosed stellar mass;  the system properties are shown for remnants both with and without a BHB.

\subsubsection{Runs with the BHB: time evolution}

In all simulations with the BHB, the merger remnant stays nearly maximally triaxial at the binary influence radius. At larger scales,  the system turns into an oblate spheroid  ($0.9\lesssim b/a\lesssim 1$) immediately after the merger is complete in most of the realizations. The evolution towards oblateness generally occurs in less than $\sim20$ time units (i.e. on a timescale of the order of the dynamical time), and cannot be a product of two-body relaxation. All the remnants in this suite of simulations are flattened, with the shorter  axis ratio $c/a$ in the range $0.6-0.7$. 

The shortest axis $c/a$ is typically aligned with the spin direction of the merger remnant, which always coincides with the spin direction of the initial galaxy merger, i.e. the positive $z$ axis. The BHB co-rotates with the galaxy (i.e. its spin is also aligned with the positive $z$ axis) in all but two runs: in run LC9b the BHB is counter-rotating, as its spin points towards the negative $z$ axis; the same happens at $t\approx t_f$ in run MC9b. However, in the latter run the  angle between the positive $z$ axis and the BHB spin progressively changes from 180 degrees to about 100 degrees at  $t=t_f+1,500$. These results are not surprising: simulations by \citet{Wang2014} already found that a BHB may form with angular momentum misaligned to the spin of the host system; in addition, \citet{Gualandris2012b} showed that BHBs whose angular momentum is initially misaligned to that of the stellar environment generally tend to realign, and this is probably what is happening in run MC9b. \\

%
%

The very large-scale structure of the system, i.e. the region including  75\% of the total mass, exhibits some oscillations over time, except for run HC5b. In this peripheral region the dynamical time over which the system finds a stable configuration is generally long due to the low stellar density, that results in a longer dynamical time: in particular, $b/a$ increases from about $0.8-0.9$ to unity in almost all runs, and by the end of the simulation the large scale system  tends to be an oblate spheroid. Again, this evolution cannot be attributed to two-body relaxation but rather to the merger itself, as the relaxation timescale in the peripheral regions of the remnant is  $\sim2\times10^6$ time units (see Table~\ref{tab:rel}), much longer than the simulated time.

The eccentricity seems to play a major role in determining the temporal evolution of the system right after the merger: on the one hand, if the initial eccentricity is small ($e\approx 0.5$) the system does not show any appreciable oscillations in the axis ratios and it reaches its equilibrium state very quickly. On the other hand, if the eccentricity is high ($e=0.9$) the axis ratios oscillate in time, and this is particularly true if one looks at the mid- or large-scale structure in systems with initial $\gamma=0.5,1$; the oscillations are related to the fact that the galactic collision occurs nearly head-on. Since the process is particularly violent, the whole system takes some time to settle down to a stable configuration; in runs LC7b, LC9b and MC9b the large-scale oscillations are still present more than 1,000 time units after the merger, even if they manifest some damping over time. The large-scale oscillations are more prominent if the system is shallower.

The persistence of the system shape also depends on the steepness of the progenitor galaxies. If $\gamma=1.5$, the remnant is really compact and its initial shape is hardly modified, even by very eccentric mergers. As a consequence, highly concentrated models reach their final equilibrium in a short time and immediately turn into oblate spheroids, with $b/a\approx 1$. When the progenitors have shallower profiles, they are more affected by the merger and, if $e>0.5$, the remnant displays some degree of triaxiality  within 25\% of the enclosed stellar mass. 

In summary, highly concentrated ($\gamma=1.5$) galaxies hosting MBHs and colliding on mildly eccentric orbits ($e=0.5$) lead to stable values for the remnant axis ratios and the resulting system is oblate, with $b/a\approx1$; shallow models ($\gamma=0.5$) on very eccentric orbits ($e=0.9$) generate mildly oblate or triaxial remnants, and exhibit strong oscillations in the axis ratios over a long timescale.  Simulations with $e=0.7$ and $\gamma=1$ show a transition behaviour between the two extremes discussed above.  We stress that none of the remnants hosting a BHB shows any degree of prolateness outside the BHB sphere of influence after the merger process is completed.

\subsubsection{Runs without the BHB: time evolution}

At very large scales  (i.e. beyond the half-mass radius) the models are unaffected by the MBHs presence, and the behaviour of the axis ratios and triaxiality parameter $T$ in the peripheral regions  is almost the same for runs with and without MBHs;  in particular, the mid- and large-scale oscillations in the shape of the system  when $e\sim0.9$ and $\gamma\lesssim 1$ are a common feature of both configurations.

At smaller scales (enclosing 1\% to 25\% of the total stellar mass) the differences between runs with and without a BHB start to be evident: if the MBHs are not included, the systems are initially nearly prolate or maximally triaxial, at least within the $\sim 10\%$ of the enclosed stellar mass if the merger eccentricity is higher than 0.5. The shortest axis of the system is generally aligned with the normal to the merger plane even in these runs without a BHB.

\subsubsection{Spurious relaxation effects}

\begin{figure}
\center
\includegraphics[trim={.0cm 0cm 8cm 18cm},width=.95\columnwidth]{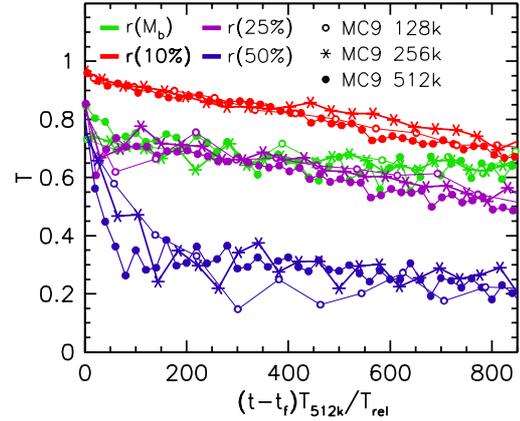}

\caption{Temporal evolution of the triaxiality parameter $T$ in the run MC9. Different lines refer to different  values of the enclosed mass (the colour code is the same as in Figure~\ref{fig:triax_t}) using a different number of particles $N$ for the simulation: $N=128k$ (empty circles), $N=256k$ (asterisks) and $N=512k$ (filled circles). The triaxiality parameter is plotted against   $(t-t_f)\cdot T_{512k}/T_{\rm rel}$, where $T_{\rm rel}$ is the local relaxation timescale   for each different model computed at $t\approx t_f$; $T_{512k}$ is the local relaxation time for the run with $N=512k$ particles. The  factor $T_{512k}/T_{\rm rel}$ ensures that  the shape evolution of different merger remnants is evaluated along the same  fraction of $T_{\rm rel}$. Lines with the same colour (i.e. evaluating $T$ on the same spatial scale) well overlap regardless of $N$; this clearly suggests that relaxation is the main driver beyond shape evolution in runs without BHB. 
}
\label{fig:Nscal} 
\end{figure}

All the runs without BHB exhibit a slow but steady growth of the axis ratios (especially $b/a$) towards unity: 
within the $\sim 25\%$ of the enclosed mass, all the remnants generally evolve towards a more oblate shape, perhaps even towards sphericity. This is due to spurious relaxation rather than to the merger process, as the shape evolution is faster for more concentrated models, i.e. when the relaxation time is shorter (Table~\ref{tab:rel}). To verify this, we re-ran simulation MC9 including a smaller number of particles, i.e. $N=128k$ and $256k$, and we compared the results with the reference simulation with $N=512k$. 
 In this comparison, we find that  models with lower $N$ systematically  evolve faster towards oblateness (i.e. lower $T$) at {\em all scales}, indicating that relaxation is  the driver behind shape evolution. In order to confirm this, in Figure~\ref{fig:Nscal} we plot the evolution  of the triaxiality parameter $T$ for different $N$, as a function of the time-related quantity

\begin{equation}
\tau=(t-t_f)\frac{T_{512k}}{T_{\rm rel}};
\end{equation}
here $T_{\rm rel}$ is the local relaxation timescale of each model (eq. \ref{eq:trel}) evaluated at $t\approx t_f$, while $T_{512k}$ is the local relaxation time for the run with $N=512k$ particles. The quantity $\tau$ coincides with  $t-t_f$ in the model with $512k$ particles, while it represents $t-t_f$ extended by a factor proportional to $T_{\rm rel}^{-1}$ in the other runs. In this way,  Figure~\ref{fig:Nscal} shows the shape evolution, for each given spatial scale, across a fixed interval of the relaxation time. In the plot, lines describing the behaviour of $T$ for a given fraction of enclosed mass (i.e., lines with the same colour) well overlap irrespective of $N$;  this is strong evidence that two-body relaxation is the main driver of the shape evolution of models without BHB.

 Obviously, relaxation effects are at play even in runs with the BHB. 
We checked this aspect by re-running simulation MC9b with lower $N$ values. The impact of relaxation is comparable to what we find in the run without any MBH (i.e., the system evolves faster towards unitary axis ratios if $N$ is lower) at scales enclosing more than $\approx 5\%$ of the total stellar mass. However, BHB hosts with high $N$ tend to have axis ratios closer to unity  within $\sim 2\%$ of enclosed stellar mass, compared to remnants with lower $N$; i.e., the trend with $N$ is inverted at small scales, compared to what is found at larger separations. This  might mean that real BHB hosts -- typically with  $N \gg 512k$ -- display slightly larger axis ratios  and lower $T$ in their inner regions, compared to what we find in our reference runs with $N=512k$. Thus the value of  $T$  within a few per cent of the total stellar mass in our runs including the BHB might be slightly overestimated, compared to real galaxies\footnote{ 
The described effect is likely rather small: the run with $N=512k$  has axis ratios larger by  $\lesssim 0.075$ compared to the axis ratios computed with $N=256k$ even at the smallest scale we consider (enclosing 0.5\% of the stellar mass), where the low-$N$ effects are most extreme; this translates in a triaxiality parameter that is smaller in our reference run by  0.05 at most, compared to the run using $N=256k$.
}. 

Given all these facts, and since real galaxies are generally unaffected by two-body relaxation, the actual shape of the merger remnants in our simulations has to be evaluated after the merger is completed, but before two-body relaxation has played a significant role in remodelling the systems.  Isolating the action of two-body relaxation from the effects of the merger is not a trivial task; 
in fact, on the one hand one needs to evaluate the shape of the remnants early enough to avoid spurious relaxation effects; on the other hand, the system settles on a stable configuration after the merger over a timescale of the order of the dynamical time. Such time interval in the outer regions of the remnants can be very long, even larger than the nuclear relaxation timescale in the same system. For this,  by considering the relaxation time as a function of radius and from Figure~\ref{fig:triax_t}, we find it best to evaluate the shape of the model within $25\%$ of enclosed mass at a time equal to: (i) $t_f+300$ if $\gamma=0.5$, (ii) $t_f+150$ if $\gamma=1$, (iii) $t_f+40$ if $\gamma=1.5$.
The shape of the system at the half-mass radius and beyond is always evaluated  at  time $t_f+500$.  Such times are used for estimating both the remnant geometry and the kinematical properties of the merger remnants, i.e. the quantities shown in Figures \ref{fig:triax_m}, \ref{fig:triax_e}-\ref{fig:beta}. The reference times are marked with vertical dashed lines in Figure~\ref{fig:triax_t}.

\subsubsection{Effect of the BHB}

The most striking result of our simulation is particularly evident from Figure~\ref{fig:triax_m}: remnants hosting MBHs always keep a lower value of $T$ at all scales compared to remnants resulting from equivalent runs without the BHB. When the massive bodies are present, the system geometry is more oblate at all scales and possibly closer to spherical. In particular $b/a$ can be as low as $0.6$ in runs without MBHs, while it is always higher than $0.8$ (and even 0.9 outside  1\% of the enclosed mass) if the BHB is present. In addition, when the MBHs are included, the system is generally less flattened (i.e. $c/a$ is higher). The differences between remnants with and without MBHs are extremely evident when the initial galaxies are less concentrated and the 
merger is more radial, while dense 
models colliding on more circular orbits (as HC5 and HC5b)
do not show any clear difference in shape. 
{A qualitative inspection of the remnants shape in time suggests that all our systems rotate; in particular,} if one excludes run HC5, runs without a BHB host a rotating bar-like structure {  aligned to the merger plane, }that may extend as far as the half-mass radius; the structure has an increasing extent and prolateness in remnants resulting from low-angular-momentum collisions of shallow systems. Outside the limiting radius, the bar-like structure loses coherence and the remnant rotates with a non-uniform angular speed. A maximally triaxial or slightly oblate figure is also present within the half mass radius in the most eccentric runs including the BHB. In all the other remnants hosting the MBHs triaxiality is mostly limited to the binary's influence radius.  In addition, Figure~\ref{fig:triax_m} shows very clearly   that $T$ never attains values higher than 0.5 outside the binary influence radius when the BHB is present.

\subsubsection{Only one MBH}
 At this stage it is worth investigating whether a BHB is necessary for producing the aforementioned differences, or even the presence of a single MBH drives the remnant towards oblateness. For this purpose, we analize simulation MC7o: such simulation has the same initial orbit and density profile as in runs MC7, MC7b but it hosts a MBH in only one of the two colliding galaxies.
The results of this comparison are shown in the central panels of Fig. \ref{fig:triax_m}. Clearly, even a single MBH erases all the triaxiality outside the MBH's sphere of influence, meaning that the differences in the shape of the remnant are not driven by slingshot ejections but probably by the steep central potential induced by the MBH(s) presence.

{Interestingly, $T$ has almost the same dependence on the enclosed stellar mass when only one or two MBHs are present; however,  the axis ratios $b/a$ and especially $c/a$ in runs MC7b and MC7o attain a different value within the MBH(s) sphere of influence: the system gets closer to spherical ($b/a\approx 0.9$, $c/a\approx 0.8$) immediately after the merger when only one MBH is present, while it is more flattened ($b/a\approx 0.85$, $c/a\approx 0.6$) when the remnant hosts a BHB.

In order to understand whether the triaxiality within the BHB influence radius survives after the BHB coalescence, we manually merged the two MBHs into one in simulation MC7b at $t=t_f+400$, and we studied the further evolution of the remnant morphology within the single MBH sphere of influence. When two MBHs are replaced with one, the inner regions of the remnant slowly migrate towards a more isotropic configuration and the axis ratios reach the same values as in run MC7o. This change of the system geometry takes roughly 500 time units to be completed, a timescale close to the relaxation time of the system at such scale, suggesting that two-body relaxation is the main driver behind the shape evolution. We thus expect the very central region of a galaxy to `remember' the presence of a BHB for a timescale of the order of its relaxation time. }

\subsubsection{Dependence on the initial conditions}

\begin{figure*}
\includegraphics[trim={.5cm 0cm 4cm 13.4cm},width=\columnwidth]{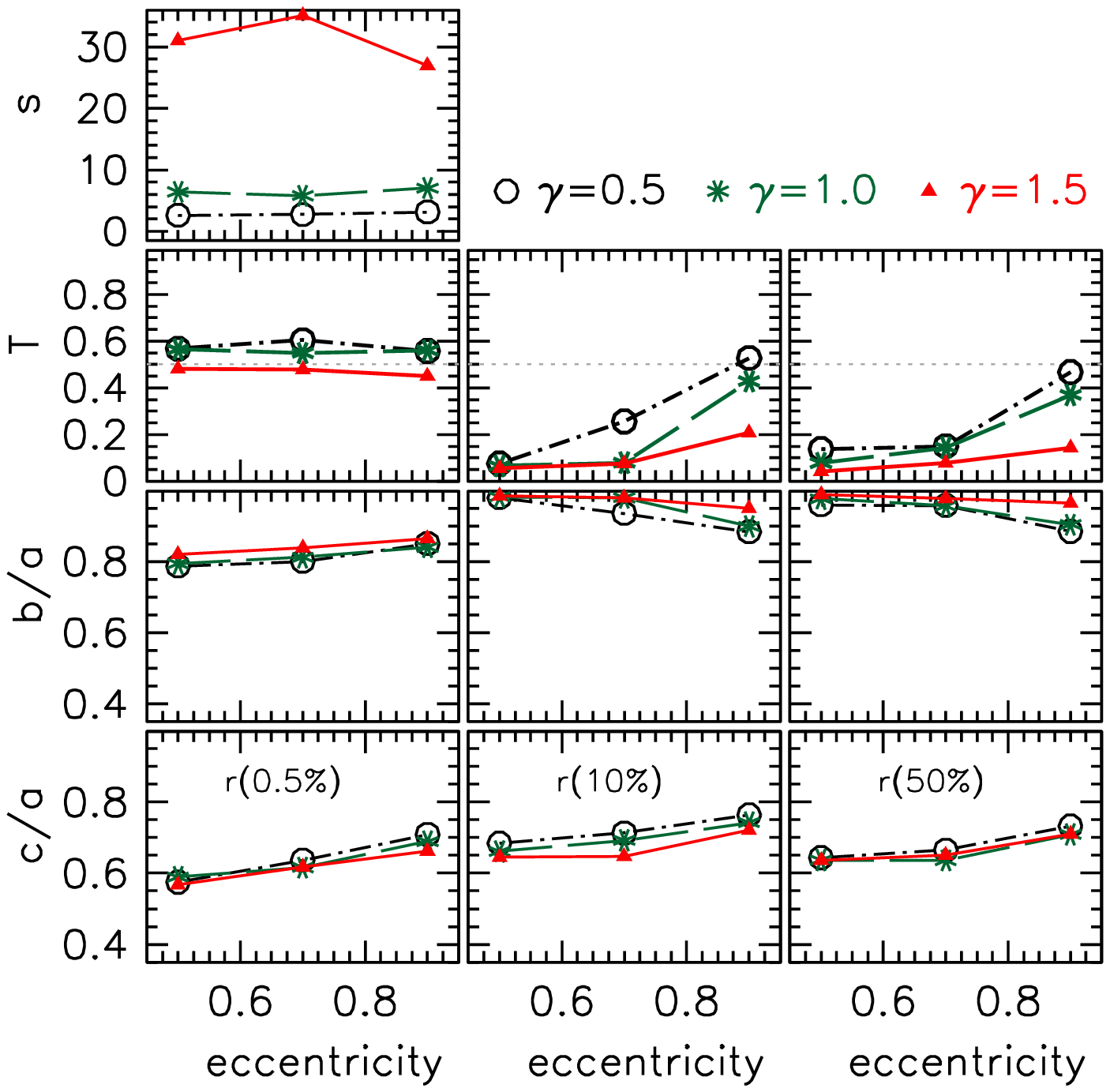} 
\includegraphics[trim={.5cm 0cm 4cm 13.4cm},width=\columnwidth]{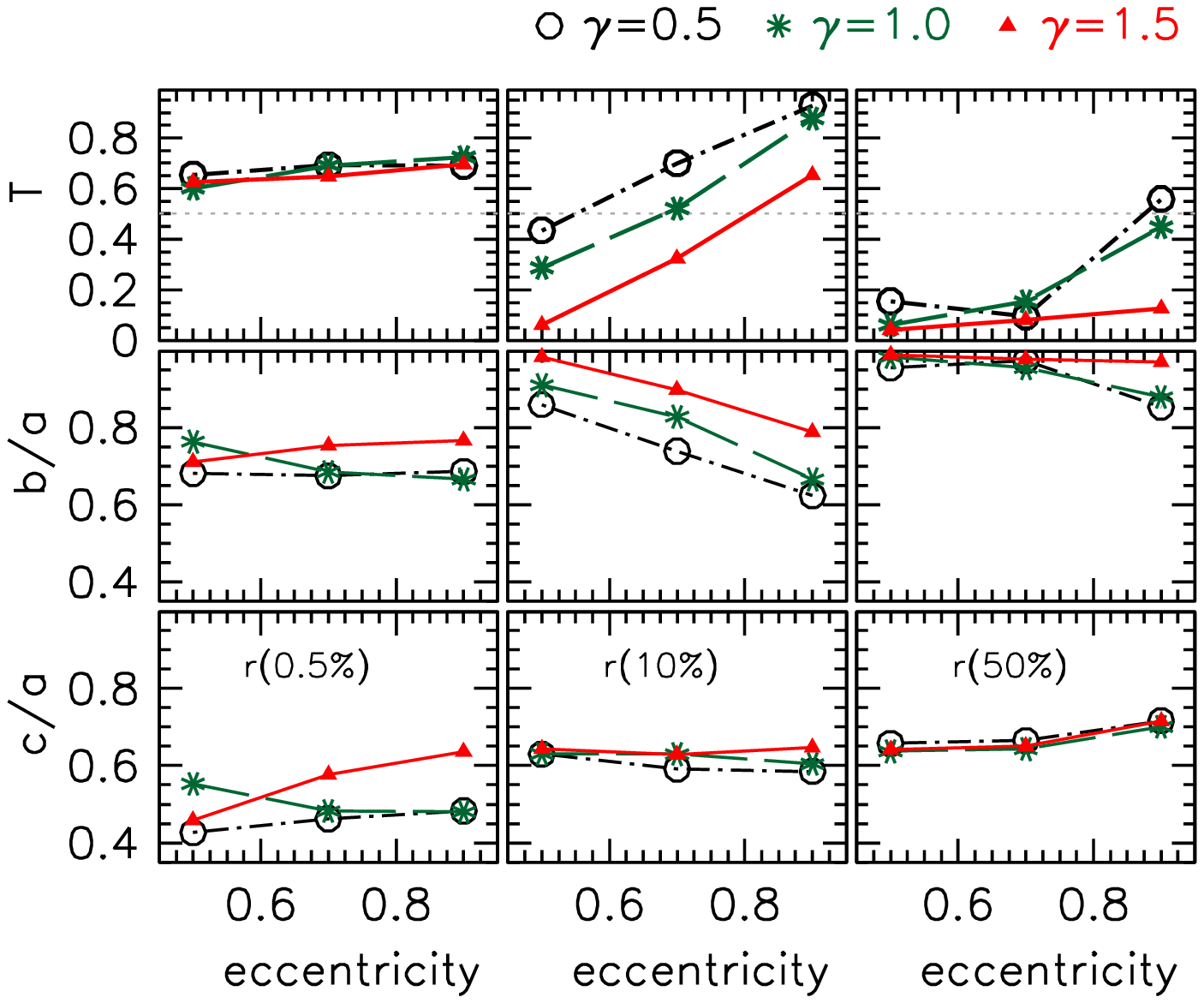}

\caption{Triaxiality parameter $T$ and axis ratios  as a function of the initial orbital eccentricity for runs with BHB (left) and without BHB (right). 
The panels show  $T$, $b/a$  and $c/a$ as a function of $e$. The upper left plot also shows the hardening rate $s$ for runs with the BHB; $s$ is  computed over the same interval of time used for  computing the triaxiality and axis ratios. The columns refer to different fractions of enclosed mass: 0.5\% (first column), 10\% (second column) and 50\% (third column). Different symbols show simulations with inner density slope of the progenitors, $\gamma$, equal to $0.5$ (black circles), $1$ (green asterisks) and $1.5$ (red triangles).}
\label{fig:triax_e} 
\end{figure*}

\begin{figure*}
\includegraphics[trim={.5cm 0cm 4cm 13.4cm},width=\columnwidth]{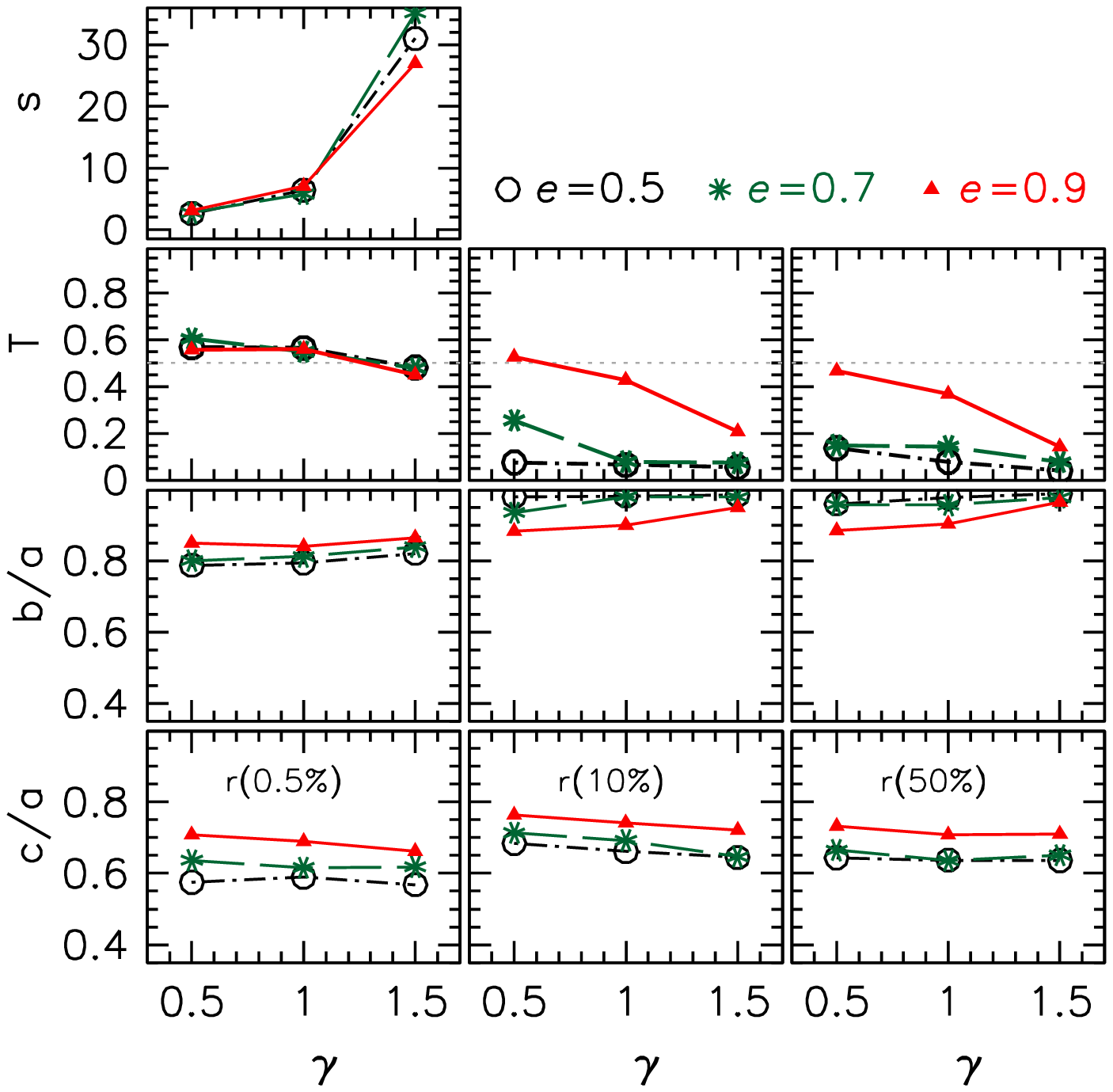} 
\includegraphics[trim={.5cm 0cm 4cm 13.4cm},width=\columnwidth]{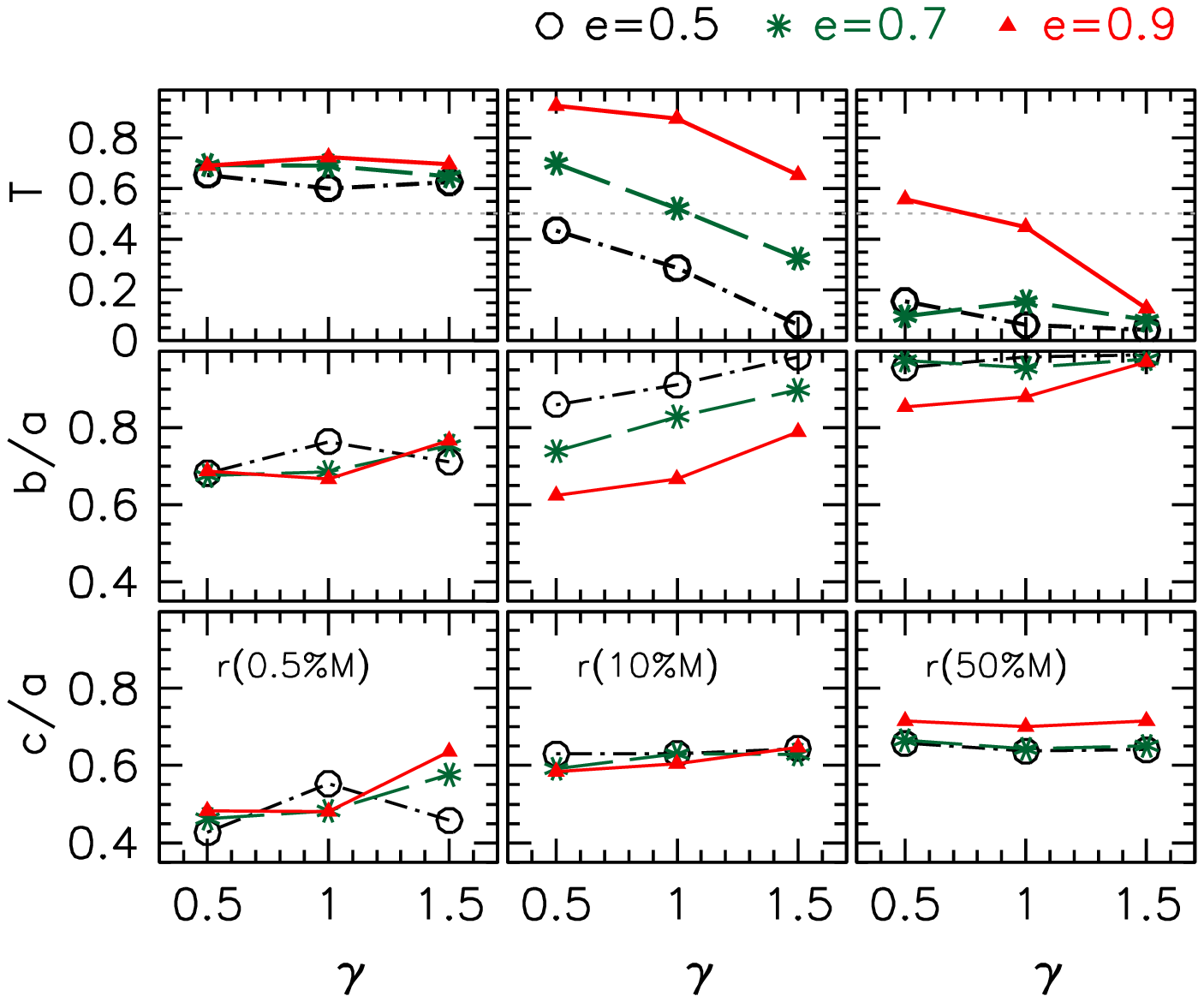}

\caption{Triaxiality parameter $T$ and axis ratios  as a function of the inner density slope of the progenitor galaxies, $\gamma$, for runs with BHB (left) and without BHB (right). 
The panels show  $T$, $b/a$  and $c/a$ as a function of $\gamma$. The  upper left plot also shows the hardening rate $s$ for runs with the BHB; $s$ is  computed over the same interval of time used for  computing the triaxiality and axis ratios.  The columns refer to different fractions of enclosed mass: 0.5\% (first column), 10\% (second column) and 50\% (third column). Different  symbols show simulations with orbital eccentricity of the merger, $e$, equal to $0.5$ (black circles), $0.7$ (green asterisks) and $0.9$ (red triangles).}
\label{fig:triax_g} 
\end{figure*}


In this section we  address how the initial conditions of the merger influence the morphology of the remnant. Figures~\ref{fig:triax_e} and \ref{fig:triax_g} show the axis ratios and $T$ dependence respectively on the merger eccentricity and on the density slope of the merging systems, for runs with and without BHB.

The axis ratio $b/a$ outside the $\sim 1\%$ of enclosed mass tends to decrease with increasing $e$, especially within the $10-25\%$ of enclosed stellar mass; this results in a higher value of $T$ when the merger is more radial, and is true regardless of the presence of the BHB. The trend in $b/a$ and $T$ is more prominent when the BHB is omitted, as the system can attain a value of $T$ that is greater than 0.6. 
When the BHB is included, an increasing eccentricity also determines the increase of $c/a$, i.e. the model is more flattened if the initial galaxies are on a higher angular momentum orbit. When the BHB is not included, this trend is clear only at large radii, while the mid- and small-scale structure exhibit a more stochastic 
 trend in $e$. At very small scales (i.e. enclosing $\sim 0.5\%$ of the stellar mass)  {$T$ seems not to depend on $e$, or the dependency is too weak to be distinguished from statistical noise.} 

The shape of the remnant also depends on the density profile of the merging galaxies: $b/a$ increases with  $\gamma$ at  any scale and independently of the presence of the two MBHs, and this results in a declining value of the triaxiality parameter $T$. Such trend is more evident at radii enclosing 10\% of the stellar mass when the BHB is omitted. The dependence  of $c/a$ on the concentration of the progenitors is less obvious: $c/a$ seems to  decline if $\gamma$ is increased in runs with the BHB, especially in the central regions of the model; the trend seems to be opposite when the remnant does not host any MBH; {however such dependencies of $c/a$ are very weak and might be a result of statistical noise.}

Figures~\ref{fig:triax_e} and \ref{fig:triax_g} also show the dependence of the hardening rate on $e$ and $\gamma$. As already mentioned, $s$ does not show any obvious relation with $e$  while it strongly increases with increasing $\gamma$; moreover, {\it there is no clear correlation between $s$ and the morphology of the system at any scale}. 

\subsection{Kinematics of the remnants}
\label{sec:vel}
\subsubsection{Rotational support}

\begin{figure}
\center
\includegraphics[trim={.0cm 0cm 4.5cm 15.5cm},width=\columnwidth]{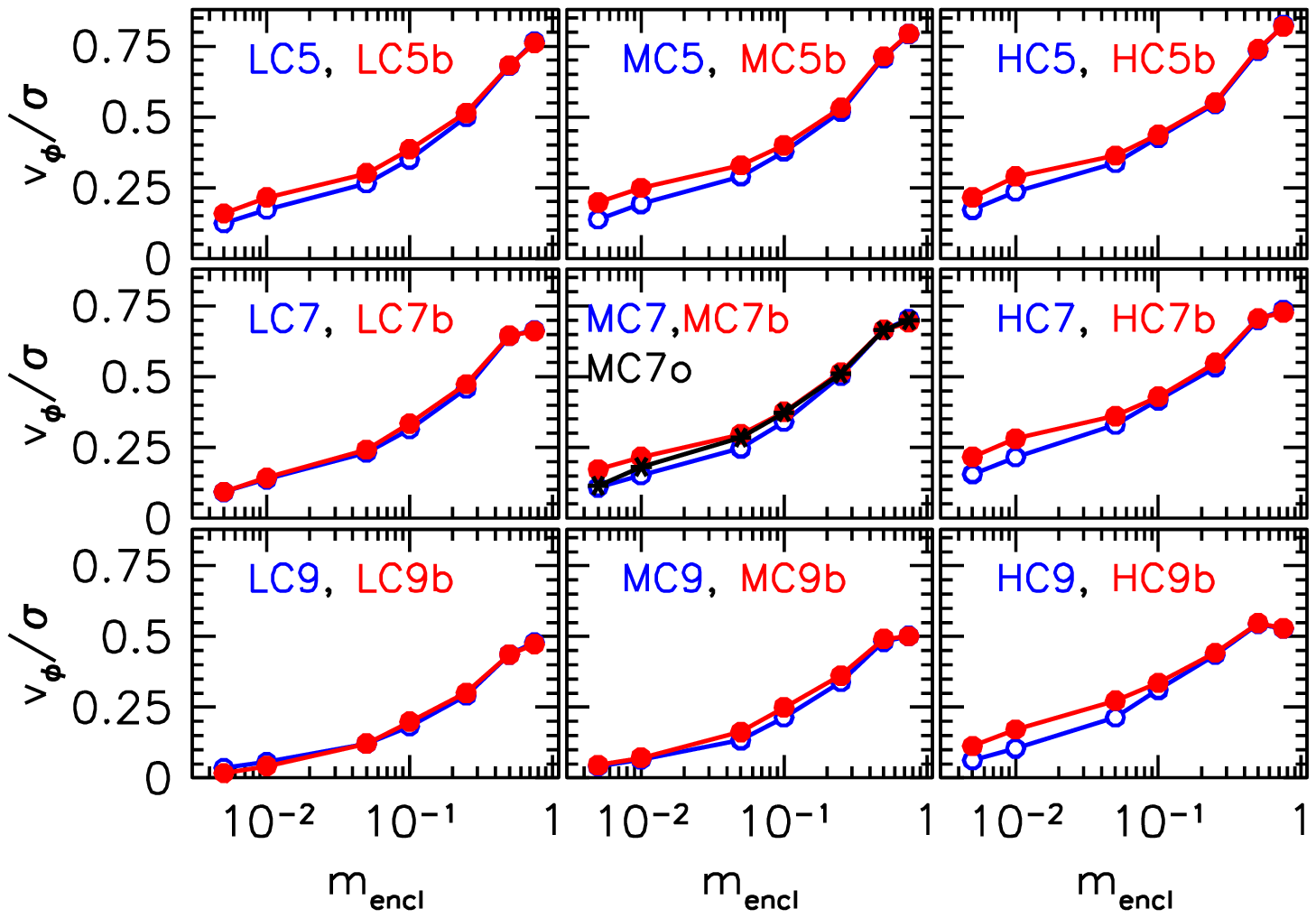}
\caption{Fraction of residual rotational velocity $v_{\phi}$ to velocity dispersion $\sigma$ of the remnant as a function of the enclosed mass $m_{\rm encl}$. Each panel is labelled with the name of the runs shown. Filled red points: runs with the BHB; empty blue points: runs without the BHB; { black asterisks: run MC7o with a single MBH}.}
\label{fig:vz} 
\end{figure}

Even if the progenitor galaxies are pressure supported systems, the merger induces a certain degree of net rotation in the remnant, whose velocity vector always lies along the merger plane.
  To quantify this, we evaluated the magnitude of the   velocity  component aligned to the merger plane\footnote{ In the computation of $v_\phi$ 
 we averaged the tangential components of the velocities, and not their magnitudes. Stars with positive and negative tangential velocities cancel each other out. Therefore $v_\phi$  describes  the degree of rotation of the system.} { ($v_{\phi}$)} and the local velocity dispersion of the remnant { ($\sigma$, computed for different enclosed masses)}. In Figure \ref{fig:vz} we show the ratio between such velocities,  $v_{\phi}/\sigma$, as a function of the enclosed mass: 
 more radial mergers  result in  less rotationally supported remnants, as expected by the laws of conservation of angular momentum;  { this is true at all radii.} The system exhibits a higher rotational support  at large distances from the centre, as most of the orbital angular momentum is absorbed by the peripheral regions in the initial phases of the merger.
{Our simulations also show that the merger relic is only partially rotationally supported: $v_{\phi}/\sigma$ can be as high as $0.8$ beyond the radius enclosing 25\% of the mass; the rotational support gradually drops moving inwards, reaching $v_{\phi}/\sigma=0-0.3$  within the 1\% of enclosed stellar mass. 

Figure~\ref{fig:vz} shows that the rotational support within 1\% of enclosed mass in most runs with the BHB is slightly higher compared to runs without MBHs and even  to the run with only one MBH. To explain this, we have to keep in mind that when a BHB is present, it expels stars on radial orbits, thus only stars on almost circular orbits can remain in the innermost regions. As a consequence,  BHB-hosts attain a higher value of  $v_\phi/\sigma$  in the centre of the remnant, while the same does not apply when only one or no MBHs are present. This effect is enhanced in highly concentrated models, as more stars can interact with the BHB.

At larger scales, including up to 10\% of the total mass, $v_\phi/\sigma$ is still higher in runs with the BHB compared to runs with no MBHs; however  the run with a single MBH behaves as the case  with the BHB, suggesting that a process different than slingshot ejection is at play. We propose such additional process to be the deposition of angular momentum due to the infall of the MBH(s), when they are present. In order to test this possibility, we adiabatically grow  a MBH of mass $0.005M_{\rm tot}$ in the remnant of the originally MBH-free run MC7, starting at $t=t_f+10$, i.e. after the completion of the merger process; the MBH is bound to grow linearly with time, reaching its final mass in 50 time units. In this test case, $v_\phi/\sigma$ behaves exactly as in run MC7 with no MBHs. This confirms that if one or two MBHs participate the merger process, their angular momentum loss due to dynamical friction increases the rotational support in the host galaxy; such effect is very small, and it is maximum at radii enclosing $5-10\%$ of the total stellar mass.}

\subsubsection{Velocity anisotropy}


\begin{figure}
\center
\includegraphics[trim={.0cm 0cm 4.5cm 15.5cm},width=\columnwidth]{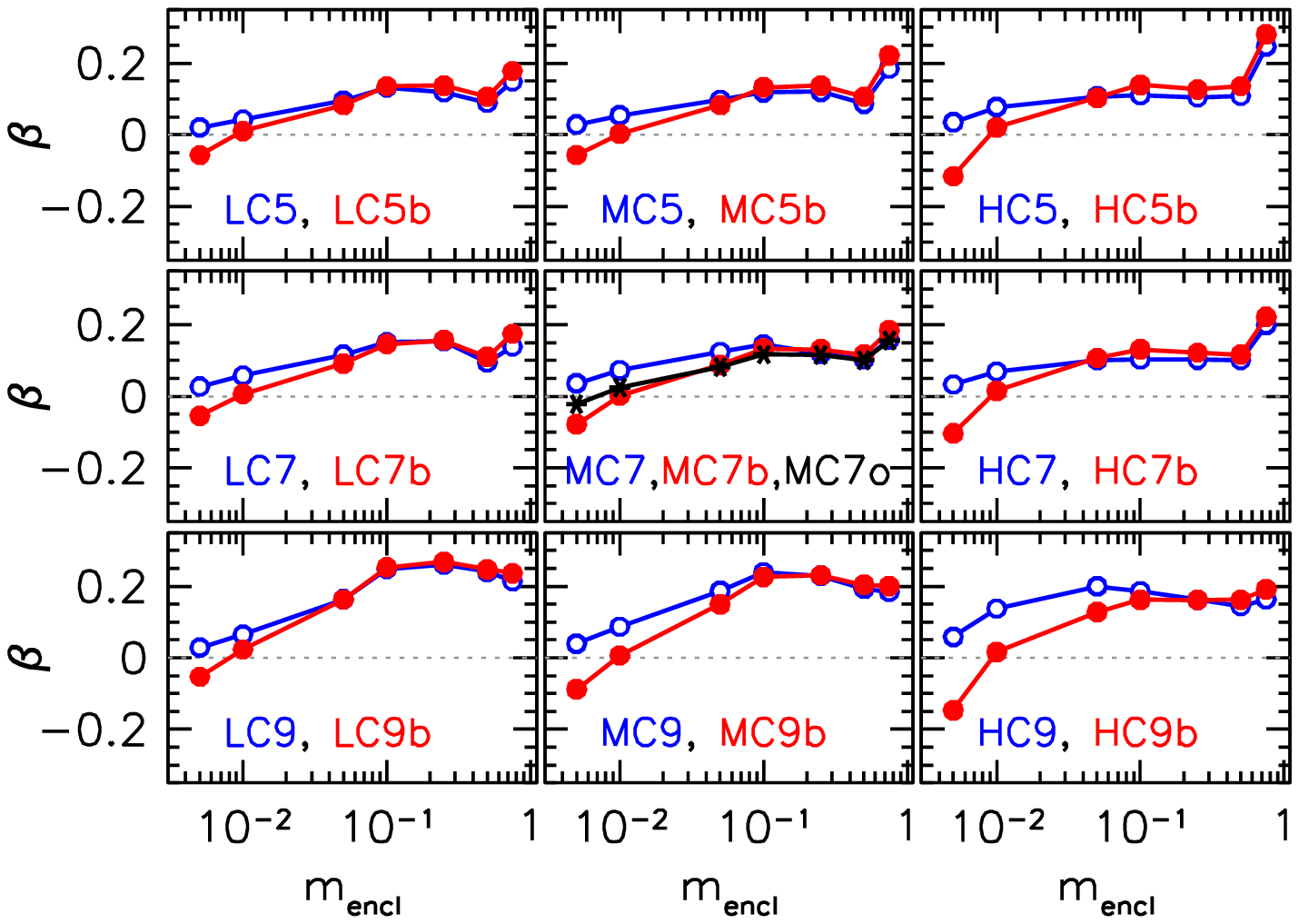}
\caption{{Anisotropy parameter $\beta$ as a function of the enclosed mass $m_{\rm encl}$. Each panel is labelled with the name of the runs shown. Filled red points: runs with the BHB; empty blue points: runs without the BHB;  black asterisks: run MC7o with a single MBH}.}
\label{fig:beta} 
\end{figure}

A stellar system can also be characterized by the \emph{anisotropy parameter} 
\begin{equation}
\beta=1-\frac{\sigma_T^2}{2\sigma_R^2},
\end{equation}
where $\sigma_T=(\sigma_{\theta}^2+\sigma_{\phi}^2)^{1/2}$ is the tangential velocity dispersion, and $\sigma_R$ represents the radial velocity dispersion. 
The anisotropy parameter measures whether a system is dominated by stars on radial orbits ($0<\beta<1$), tangential orbits ($-1<\beta<0$) or the two are perfectly balanced and the system is isotropic ($\beta=0$, as in the  progenitor galaxies).

{The anisotropy parameter as a function of the enclosed stellar mass  is shown in Fig. \ref{fig:beta}: in all runs, stars are mainly found on radial orbits beyond 1\% of the enclosed mass, as $\beta$ mostly lies in the range $0.1-0.3$; generally $\beta$ attains higher values if the galactic merger is more radial, at least within the half-mass radius. Runs including the BHB exhibit a lower value of the anisotropy parameter at small scales; in particular, within a sphere including 0.5\% of the mass $\beta$ stays between 0 and 0.05  if the BHB is absent, while  it lies in the range $[-0.15,-0.05]$ if the BHB is included. Such behaviour is again easily explained in terms of BHB slingshot ejections: at small scales, the BHB ejects  stars on radial  orbits, allowing only stars on tangential orbits to survive in the inner regions of the system. Such BHB-induced small scale effect  is enhanced in runs with a high central concentration.

The anisotropy parameter for run MC7o with a single MBH  is also shown in Fig. \ref{fig:beta}: 
at small scales, the system is less tangentially biased compared to the run with a BHB, as slingshot interactions are not at play; however angular momentum deposition due to the infalling MBH  enhances the tangential anisotropy in this run compared to the run with no MBHs, as already mentioned in the previous section.

Finally, in the peripheral region of the remnants, runs with the BHB are more radially biased compared to runs with one or no MBHs: this is the large scale effect of slingshot interactions that scatter stars on very radial orbits.  }

\subsection{Fraction of escapers}

\begin{figure}
\center
\includegraphics[trim={1cm 1cm 1cm 13.5cm},width=\columnwidth]{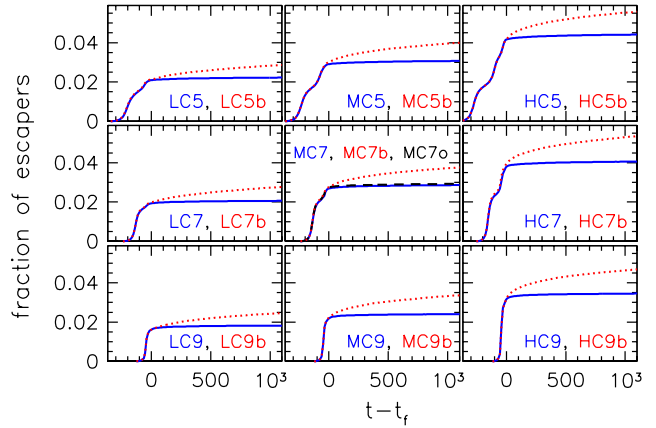}
\caption{The different panels show the fraction  of stellar escapers as a function of time in runs with the BHB (red dotted lines), with only one MBH (run MC7o, black dashed line in the central panel) and without MBHs (blue solid lines). Each panel is labelled with the name of the runs shown. }
\label{fig:esc} 
\end{figure}

The fraction of stellar escapers as a function of time is shown in Figure~\ref{fig:esc}. During the merger, escapers are mainly produced after each pericentre passage: in this stage bound stars get destabilized due to the strong perturbation in the global potential, and may leave the system. As a consequence, more escapers are produced within $t_f$ when the orbit is less eccentric: the merger is slower and the two galaxies undergo multiple pericentre passages before reaching coalescence. In addition, stars in more concentrated systems are more efficient at producing escapers: runs with $\gamma=1.5$ generally have a fraction of escapers at $t_f$ that is about two times the fraction of escapers in runs with $\gamma=0.5$. Such effect is predicted by the theory of violent relaxation \citep{Lynden-Bell1967}: if the progenitor systems are cuspier, stars within each galaxy feel a stronger variation in the total gravitational potential during the merger, thus the average change in the energy per unit mass of each star is expected to be more substantial, and the fraction of stars gaining sufficient energy to escape the system is greater.

If the system does not host a BHB, escapers are no longer produced after the merger process is completed. In contrast, when a BHB is present a number of stars undergo slingshot interactions and get ejected from the remnant after the BHB semimajor axis has dropped below $a_h$. The effect of the BHB is clearly visible in Figure~\ref{fig:esc}: the fraction of escapers steadily increases after $t_f$ in remnants harbouring a BHB. More escapers are produced by the slingshot interactions if the initial system is more compact, as more stars are initially available on low energy orbits: the BHB  generally unbinds two times more stars if $\gamma=1.5$ compared to runs with $\gamma=0.5$. 
When only one MBH is included in the simulation, the number of escapers grows almost as it does in the analogous run without MBHs (Figure~\ref{fig:esc}, central panel), confirming that  BHB slingshot ejections determine the continuous production of escapers after $t_f$.

\section{Summary and discussion}\label{sec:disc}
In this paper we carried out a suite of equal mass galaxy merger $N$-body simulations, varying the initial orbit and inner density slope of the merging galaxies and including or not a MBH in the centre of the colliding systems; when a MBH is included in each merging galaxy, a BHB forms in the centre of the remnant.  {   Using convergence tests and analytic estimates for the two-body relaxation time, we minimised the effect of spurious two-body relaxation by analysing our simulations at a time and on a spatial scale at which two-body relaxation time is always  longer than the simulation time.}

Our main aim was to analyze the link between the morphology and kinematics of the newly formed stellar system and:  (i) the initial orbit and density profile of the two progenitor galaxies and (ii) the presence or absence of a BHB (or even a single MBH) in the centre of the merger relic; finally, we studied how the shape of the remnant influences the BHB hardening efficiency.
In what follows we summarize and discuss our main findings.

\subsection{Morphology of the remnant and merger initial conditions}

As expected, the mid- and large-scale geometry of the system strongly depends on the merger orbit: high angular momentum collisions generally lead to the formation of an oblate spheroid, while radial (or equivalently, lower impact parameter) mergers can produce maximally triaxial or (if no MBHs are present) prolate systems, in agreement with the findings in \citet{Gonzalez2005b}. This might be linked to the fact that the initial conditions for more radial galaxy collisions are more `anisotropic' (i.e. the orbit of the galaxies is elongated in the merger plane, instead of being close to circular), thus the projection of the final remnant in the  merger plane is more stretched. {  Alternatively, this effect may be a result of radial orbital instability, that drives a break in the symmetry of the system \citep{Antonov1987,Saha1991, MacMillan2006, Barnes2009}.} 
Higher angular momentum mergers were also found to produce more flattened remnants (especially beyond the half-mass radius): as expected,  a higher degree of net rotation is induced in the outskirts of the system for more circular collisions, and  in turn  the remnant  shape appears to be more flattened.

The shape of the merger relic is also connected to the galaxy progenitors'  density profile: while collisions between more concentrated galaxies generally produce oblate spheroids with $b/a$ closer to unity, low concentration systems are found to be often maximally triaxial and possibly  prolate. This may  be connected to the fact that stars in systems with a shallower density profile are more sensitive to tidal torques, thus their orbit is more easily modified during the merger: it follows that the collision between less concentrated galaxies can affect  the remnant shape more, and the resulting system will better remember the imprint of the merger orbit.

{ 

We further stress that the shortest axes ($c/a$) of the oblate systems produced in our simulations are always almost perpendicular to the merger orbital plane. In systems with the BHB, the orientation of the merger plane  almost always coincides with the orientation of the BHB orbital plane, thus one may  think that the BHB causes the system to be flattened in the direction of its angular momentum vector. However, even in runs with only one or no MBHs the system is flattened in the direction of the merger plane, suggesting instead that the galaxy collision (and perhaps the resulting rotation) influences the orientation of the principal axes of the ellipsoid.
}

\subsection{The role of the central MBHs}
Perhaps the most striking result in our  simulations is the fact that if at least one MBH is involved in the merger, the system shape is  noticeably influenced by the MBH well beyond its  sphere of influence, and this is true from the very moment the remnant forms. Starting from the same initial orbit and density profile of the merging galaxies, we found that the central regions of merger relics hosting MBHs are always closer to spherical, and the triaxiality parameter $T$ is noticeably smaller compared to the same runs without  MBHs: a merger product hosting one or two MBHs is generally found to be oblate, {  aligned with the galaxy merger plane},  and never attains $T>0.6$,  while when no MBHs are present the remnant is typically prolate (sometimes reaching $T\approx1$) or maximally triaxial. The aforementioned differences are particularly evident within a radius including approximately $2-50$ times the mass of the central MBH(s), while the remnant shape is generally the same beyond $\approx100$  MBH(s) masses.

The fact that a central massive body may render the system rounder or closer to oblate was already evident from a number of studies \citep[e.g.][]{Lake1983,Gerhard1985,MerrittQuinlan1998,Holley-Bockelmann2002}, which  addressed the evolution of equilibrium mass models (rather than merger relics) where a MBH was adiabatically grown. In these studies, the evolution of the galaxy shape is  attributed to the fact that the MBH acts as a scattering centre, rendering centrophilic orbits stochastic: the volume filled by the scattered orbits is rounder and does not support the original galaxy shape, thus the global morphology of the system changes\footnote{The final axis ratios of the system in these studies were  generally closer to unity compared to what we find here.}.

Similar studies  found that a central strong cusp may have an analogous effect: it may act as an orbit scatterer  evolving the system towards a more spherical shape  \citep{MerrittFridman1996,MerrittValluri1996}. This could be the reason why, among our models without MBHs, the ones with the highest initial concentration  were those that either became immediately oblate, or retained a shape with $T>0$  for a very short timescale.

Even if a MBH (or possibly even a strong cusp) seems to drive the system towards oblateness or sphericity, a series of more recent studies were able to demonstrate that steady triaxial ($T\leq0.5$) models involving a high fraction of chaotic orbits and hosting a MBH can be constructed, and they were found to retain their shape over many dynamical times \citep{MerrittPoon2002,MerrittPoon2004}. This means that triaxiality can be achieved in systems hosting a central MBH. To our knowledge though, a stable steady state solution for a prolate system   hosting a MBH has never been found: according to \citet{MerrittPoon2004}, mildly prolate  models harbouring a central massive body always evolve towards an oblate axisymmetric configuration, in agreement with  the fact that our MBHs-hosting remnant never reach $T>0.5$  outside the BHB sphere of influence.  To our knowledge, the fact that $T$ seems to have an  upper limit  if the system hosts a central massive body has no thorough explanation, and we reserve to better analyse this aspect in a forthcoming paper.

Recently, \citet{Vasiliev2015} studied the evolution of BHBs embedded in stable triaxial mass models via a Monte-Carlo method that allows them to switch off two-body relaxation effects: when they analyse the shape of a system hosting a BHB, they also find an  evolution towards axisymmetry, while when only a single MBH is present the morphology of the system does not change significantly over a long timescale \citep{Vasiliev2015b}. For this reason, they suggest that the shape evolution observed in studies involving a single MBH \citep[e.g.][]{MerrittQuinlan1998,Holley-Bockelmann2002}  may be greatly affected by spurious
 two-body relaxation effects \citep{Kandrup2000}; when a BHB is included though, they speculate that resonant perturbations of chaotic stellar orbits resulting from the BHB time-dependent potential \citep{Kandrup2003} may cause the observed shape evolution even if relaxation is not at play.

{  The results by \citet{Vasiliev2015} cannot be easily compared to ours,  as we form our merger remnants self-consistently from galaxy collisions, inducing some rotation in the systems. But we can state with confidence that our results, on the scale and at the times we analyse the simulations, are not affected by spurious numerical two-body relaxation.}  Thus we propose that galactic collisions may have an important role in determining the differences in the shape of remnants with and without MBHs.
 Given that evolution towards oblateness is believed to result from the scattering of stars into chaotic orbits due to the  MBH(s) presence, we propose that the merger itself may facilitate such scattering process (e.g. though violent relaxation) even if two-body relaxation is not at play, and regardless of whether one or two MBHs are present.

\subsection{One or two MBHs}

Even if the mid- and large-scale geometry of a remnant hosting only one or two MBHs is very similar, some differences arise within the MBH(s)  sphere of influence. Small-scale differences in the  geometry of galaxy centres are of great importance, as they might give interesting observational constraints for distinguishing systems that host (or hosted) a BHB from systems with a single MBH. In our remnants, when only one MBH is included, the geometry of the small-scale system is visibly rounder and less flattened compared to remnants hosting a BHB\footnote{
In the interpretation of such result, one should keep in mind that the described simulations started from simplistic initial conditions, i.e. from perfectly spherical, isotropic and non-rotating systems.  In reality,  progenitor galaxies undergoing a merger are likely to have  suffered a number of mergers in their history, thus they possibly already exhibit some degree of non-spericity and some net rotation prior to the merger. 
For this, in principle a galaxy may appear rounder in its inner parts just because it underwent a number of repeated mergers, irrespective of the presence of a BHB. Such aspect deserves a further investigation in a forthcoming paper.
}. We also found that the system keeps the more flattened shape typical of a BHB-hosting remnant for about a relaxation timescale after the BHB coalescence.

The small-scale shape differences  between systems with one or two MBHs could be related to the fact that when only a single MBH is present, the galaxies' inner cusps are not destroyed during the merger process and it is difficult to perturb their spherical shape due to their compactness. When two MBHs are present instead, the central cusp within each merging galaxy is destroyed by BHB-induced stellar scatterings and the concentration of the merger relic is noticeably lowered in the centre, thus stars are more affected by global torques and the shape of the system is more easily modified. We further note that the BHB potential is as not spherically symmetric as the single MBH potential is; the elongated and time dependent BHB potential may also affect the stellar orbits within its influence radius and render the small-scale structure of the system more triaxial.\\

\begin{figure}
\includegraphics[trim={0cm 0cm 10cm  22.3cm},width=\columnwidth]{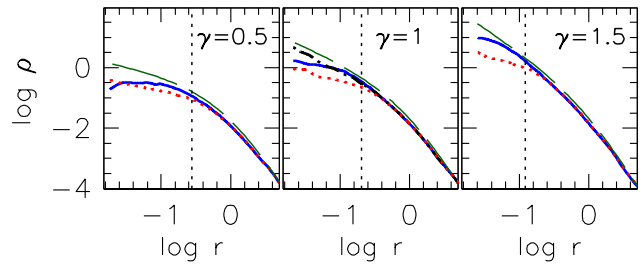} 
 \caption{Stellar mass density $\rho$ as function of the radius $r$ for merger remnants produced by a galactic collision with eccentricity $e=0.7$; the curves refer to time $t_f+250$. The progenitor galaxies' density profile is a Dehnen model with (from left to right) $\gamma=0.5, 1, 1.5$.  We show the density profile of remnants hosting a BHB  (red dotted lines) and no MBHs (blue solid lines); the black dash-dotted line in the central panel shows the density profile of run MC7o with a single MBH. For comparison, the dark-green long-dashed line shows the density profile of the progenitor galaxies; the vertical dotted line marks the radius including two times the mass of binary in runs including it.}
    \label{fig:density}
\end{figure}

There is a second effect  that may be observationally helpful in distinguishing between systems with only one or two MBHs: the so-called core scouring. If the system hosts a BHB, the binary-induced slingshot ejection of stars produces a lack of stellar mass in the central parts of the remnant. As a consequence, the inner density profile of a systems hosting (or that hosted) a BHB is expected to be less cuspy. We verified the occurrence of core-scouring in our simulations, and the results are shown in Figure~\ref{fig:density}:  we compare the density profile of the progenitor galaxies with the density profile of remnants hosting zero, one and two MBHs. When the merger relic hosts a BHB, the density profile is noticeably carved out beyond the radius including two times the BHB mass; this is  a well established result \citep[e.g.][]{Milosavljevic2001,Gualandris2012} and is supported by observations   \citep{Ferrarese1994,Lauer1995, Bonfini2018}.  
As a matter of fact, even the infall of a single MBH can produce a core in the system \citep{Gualandris2008,Goerdt2010}; however our simulations suggest that in the single MBH scenario the effect of core scouring is significantly smaller  compared to the BHB case (Figure~\ref{fig:density}, central panel). In addition, we note that the small-scale density profile of remnants not hosting any MBH is shallower compared to the one of the progenitor systems. This seems at odds with the results of \citet{Dehnen2005}, who asserts that the steeper cusp should always survive in a merger; however, the flattening of the inner density profile we observe in our simulations without MBHs is most probably an effect of numerical relaxation. We suggest the density profiles in Figure~\ref{fig:density} are more reliable outside the radius where the remnants without MBHs and their progenitors start having a comparable density profile.

\subsection{Rotation and velocity anisotropy}

In our simulations, we find that the merger induces some rotation in the final remnant, even if the bulk of angular momentum is absorbed by the outer regions of the relic, in agreement with previous findings \citep{DiMatteo2009,Gonzalez2005b}. 
As expected, our simulations show that rotational support is enhanced if the merger eccentricity is lower. 

When one or two MBHs are present in the simulation, they considerably influence the kinematics of the final remnant: when a BHB is present, it ejects stars from the core of the system, which is found to be more rotationally supported. In addition,  MBHs lose their angular momentum when sinking towards the centre of the remnant, thus they both enhance  the rotational support and lower the radial anisotropy parameter.

 Angular momentum injection, together with high central concentrations,  was previously found to irremediably change the shape of dark matter haloes  \citep{Debattista2008}, thus it might connect  with the morphology evolution we see in our runs, but we reserve to better investigate this aspect in a future paper.

Concerning the enhanced tangential anisotropy in the centre of systems hosting a BHB,  our findings are at least in qualitative agreement with the observational results of by \citet{Thomas2014}: they  find hints of kinematical tangential anisotropy in the centres of elliptical galaxies hosting a depleted core, and such depleted core may well be the fingerprint of an evolving BHB. The measurements by \citet{Thomas2014} indicate that the inner regions of elliptical cored galaxies may be even more tangentially biased compared to what we find in our runs: however, one should consider that real galaxies  (especially large ellipticals) have likely been through a number of  mergers, each of which may have contributed to boost the tangential anisotropy in the inner regions.

\subsection{BHB evolution}
We find that the evolution of the BHB is not clearly  connected with the shape of the host system, even if a clear correlation is present between the hardening rate and the concentration of the BHB host. 
In principle, such lack of connection might be due to the fact that the system looks always maximally triaxial within the BHB
influence radius, and such triaxiality might ensure a similar hardening rate in simulations with the same $\gamma$. However, \citet{Vasiliev2015}  point out that stars participating in the binary shrinking come from large 
distances from the centre, and this would mean instead that the maximum triaxiality within the BHB influence sphere does not influence the binary shrinking rate. 
%

Alternatively, the lack of connection between the system shape and the BHB hardening  could be related to the fact that all our remnants have a similar (nearly oblate) geometry at the largest scales considered in this paper. In principle, one could also consider the possibility that spurious numerical relaxation plays  the major role in determining the BHB hardening; however \citet{Vasiliev2015} recently showed that relaxation has a long-term effect on the BHB shrinking efficiency only when the host system is perfectly spherical or axisymmetric, that is never our case.
Thus we suggest that a  BHB could harden at the same rate for a given concentration of its host system, and the value of such fixed hardening rate might represent a critical value that would allow to simplify the forthcoming  studies about BHBs evolution towards GW emission. Even if we do not have a throughout  explanation for the aforementioned findings, we plan to perform an orbital analysis of our merger remnants in order to get a better understanding of the BHB hardening connection with the geometry of their hosts.

{   A further interesting result is that the BHB hardening rate tends to the same value towards the end of all our runs (see e.g. Fig. \ref{fig:s}); this perhaps reflects the fact that the BHB separation almost  reaches the softening length by the end of the simulation, and possibly slows down the BHB hardening in more concentrated models. On the other hand, the decline of $s$ in our runs is consistent or even less conspicuous compared to what found by \citet{Vasiliev2015} using a (totally different) Monte Carlo integration method; we suspect that the slowing down of $s$ we see here is thus a real effect, possibly linked with the idea that the loss cone is replenished at approximately the same rate in all models, once the initial loss cone population has been entirely scattered and the system geometry has found its equilibrium state.}

\subsection{Conclusions}

This study shows that BHBs formed from the dry merger of elliptical galaxies have a strong impact on the geometry of their host systems. In particular, binary (or even single)  MBHs render the host system more oblate, {  aligned with the orbital planes of both the BHB and the galaxy merger,} up to a radius enclosing  $\sim 100$ MBH masses,  compared to remnants produced by the merger of the same galaxies not hosting any massive body. In addition, the results of this investigation show that remnants hosting a single or binary MBH never attain  a triaxiality parameter $T>0.6$, despite merger relics not hosting any MBH generally exhibit a prolate inner figure.  
Furthermore, we find that stars within the influence radius of a single MBH are distributed in a more compact and nearly spherical geometry, while the same region appears to be  cored and triaxial if the system hosts a BHB. 

Our study points towards a possible connection between the geometry of a galactic nucleus and the presence of zero, one or two massive central bodies. {  Our findings so far qualitatively support recent observations reported in e.g. \citet{Dullo2015} and \citet{Foster2017}, but we will perform a more quantitative analysis of this, properly projecting the simulations into observables in a forthcoming paper.}


Another major finding was that the BHB shrinking rate seems to vary only with the central density of the host, while it appears to be less related to the geometry of the merger remnant. Such result might be particularly relevant for low-frequency GW science, as the timescale needed for a BHB to reach the GW-emission stage could be assumed to scale only with the core density of the merger remnant; however further studies must be carried out to pinpoint the physical reasons behind this finding.

{  Finally, our work confirms the idea that BHBs are able to reach their coalescence phase within a Hubble time in most galaxies, even if the BHB host systems are generally found to be nearly axisymmetric outside the binary influence radius.}

\section*{Acknowledgements}
We thank the anonymous  referee for their useful comments and suggestions, that improved the paper.
We also thank Eugene Vasiliev, Zoltan Haiman, Paolo Bonfini, Jean Thomas, Paolo Salucci, Mario Spera and Michela Mapelli for useful discussions and suggestions. 
We acknowledge the CINECA Award N. HP10CP8A4R and  HP10C8653N, 2016 for the availability of high performance computing resources and support. Part of the Numerical calculations have been made possible through a CINECA-INFN agreement, providing access to resources on GALILEO and MARCONI at CINECA. EB acknowledges financial support from the Istituto Nazionale di Astrofisica (INAF) through a Cycle 31st PhD grant, from the Fondazione Ing. Aldo Gini and from INAF-Osservatorio Astronomico di Arcetri through the `Stefano Magini' Prize.




\bibliography{biblio} 

\bsp	
\label{lastpage}
\end{document}